\documentclass[12pt]{article}

\newcommand{\bx}{\boldsymbol{x}}

\usepackage{algorithm}
\usepackage{algpseudocode}
\usepackage{dsfont}
\usepackage{fancyhdr} 
\usepackage{hyperref} 
\usepackage{amsmath}
\usepackage{amssymb}
\usepackage{natbib}
\usepackage{graphicx}
\usepackage{authblk}
\usepackage{booktabs}
\usepackage[margin = 2cm]{geometry}

\title{Neural Methods for Multiple Systems Estimation Models}

\author[1]{Joseph Marsh\thanks{Corresponding author}}
\author[2]{Nathan A. Judd \thanks{Contribution was carried out at the School of Mathematics, University of Birmingham, UK.}}
\author[1,3]{Lax Chan}
\author[1]{Rowland G. Seymour}

\affil[1]{School of Mathematics, University of Birmingham, UK}
\affil[2]{Okinawa Institute of Science and Technology, Japan}
\affil[3]{Dipartimento di Studi per l'Economia e l'Impresa, Università degli Studi del Piemonte Orientale, Italy}
\date{}

\begin{document}

\maketitle

\abstract{Estimating the size of hidden populations using Multiple Systems Estimation (MSE) is a critical task in quantitative sociology; however, practical application is often hindered by imperfect administrative data and computational constraints. Real-world datasets frequently suffer from censoring and missingness due to privacy concerns, while standard inference methods, such as Maximum Likelihood Estimation (MLE) and Markov chain Monte Carlo (MCMC), can become computationally intractable or fail to converge when data are sparse. To address these limitations, we propose a novel simulation-based Bayesian inference framework utilizing Neural Bayes Estimators (NBE) and Neural Posterior Estimators (NPE). These neural methods are amortized: once trained, they provide instantaneous, computationally efficient posterior estimates, making them ideal for use in secure research environments where computational resources are limited. Through extensive simulation studies, we demonstrate that neural estimators achieve accuracy comparable to MCMC while being orders of magnitude faster and robust to the convergence failures that plague traditional samplers in sparse settings. We demonstrate our method on two real-world cases estimating the prevalence of modern slavery in the UK and female drug use in North East England.}

\section{Introduction}

The estimation of population sizes for groups that are hidden or difficult to reach remains a long-standing methodological challenge in quantitative sociology. Hidden populations are groups lacking a comprehensive sampling frame due to their marginalized status, engagement in stigmatized or illicit behaviours, or lack of social visibility. Estimating the size of hidden populations present significant hurdles for conventional survey and census methods. The applications for accurate size estimates are vast and span numerous subfields, carrying important policy and theoretical implications. For instance, researchers studying people experiencing homelessness have generated estimates of the number of hospital admissions associated with people experiencing homelessness, in order to understand hospital service provision \citep{Luchenski2025}. When understanding Social Determinants of Health, accurately estimating the number of people who use illicit drugs can support intervention planning \citep{King2008}. Similarly, sociologists and demographers seeking to understand modern slavery or human trafficking rely on population size estimates to quantify the scale of human rights abuses and design effective interventions \citep{Silverman2020}. Further applications range from estimating the size of undocumented migrant populations and homeless individuals to quantifying participants in clandestine political movements, all of which are important for building accurate social theories and informing evidence-based policy. 

Multiple Systems Estimation (MSE) is a widely adopted capture-recapture technique used to estimate the size of a hidden population by comparing a set of incomplete lists, or ``systems'' that all draw from that population \citep{Bird2018}. These lists, which might include administrative records from police, hospitals, social services, or non-governmental organizations, are not required to be random samples and are typically convenience samples. The core logic of MSE rests on analysing the degree of overlap, or the number of individuals who appear on more than one list. By observing these patterns of overlap across the lists, researchers can statistically model the dependencies between the systems. This model is then used to predict the one unobserved quantity: the number of individuals who were not captured by any of the systems. The total population size is then estimated by adding this ``dark figure'' of unobserved individuals to the total number of unique individuals observed across all lists combined.

Statistically, inference for MSE models can be conducted through two primary paradigms. The traditional approach relies on maximum likelihood estimation, which provides point estimates for the unknown population size ($N$) under a specified set of model assumptions, typically within a log-linear model framework. This method seeks the single value of $N$ that makes the observed data most probable. Alternatively, inference can be conducted using Bayesian methods, which have become increasingly prevalent. This approach typically employs Markov chain Monte Carlo (MCMC) techniques to approximate the full posterior distribution of the population size. The Bayesian paradigm offers a more complete picture of uncertainty, yielding the entire posterior shape rather than relying on asymptotic confidence intervals. It is particularly valued for its ability to formally incorporate prior information and its flexibility in handling model uncertainty and complex scenarios, such as those with sparse data.

Despite the utility of MSE, its practical application in sociological research is often complicated by two significant challenges. First, the administrative and observational data sources available to sociologists are frequently imperfect. Data may be censored, where individual data and counts are suppressed to avoid identification of the individuals in the dataset. The data may also suffer from missing observations, where the full data were not able to be collected, which is often the case when working with data from conflicts or found data. These data quality issues can make it difficult to reliably estimate the list dependencies and interaction terms that are crucial for accurately predicting the unobserved population. 

Second, a growing amount of this sensitive administrative data is only accessible through secure or trusted research environments. While these environments are essential for protecting respondent confidentiality, they often impose constraints on available computing resources and software \citep{Kavianpour2022}. This computational bottleneck is particularly problematic for the Bayesian MCMC methods described earlier. While MCMC is statistically appealing for its ability to handle uncertainty, its high computational cost and long run-times can render it impractical or infeasible within the typical constraints of a trusted research environment. \citet{King21} provides the only MSE model that allows for censored counts, where maximum likelihood estimates are developed considering all possible values of the censored values. However, this can only be carried out in a frequentist way in environments with low-computational resources. 

In this paper, we propose novel Bayesian inference methods for MSE models based on two simulation-based techniques: Neural Bayes Estimators (NBE) and Neural Posterior Estimators (NPE). Simulation-based inference offers a powerful alternative to traditional methods, particularly for the complex models required to handle sociological data. Instead of relying on explicit likelihood calculations, which can become intractable or difficult to evaluate in the presence of censored or missing data, these neural methods learn to approximate posterior quantities of interest by simulating data under different parameter settings and comparing the results to observed data. A key advantage of these estimators is that they are amortised. Once the neural network is trained, a computationally intensive step that can be performed ``offline'' on unrestricted hardware, it can generate posterior estimates almost instantly for any given dataset. This ``train once, infer many times'' capability stands in contrast to MCMC, which requires a new, costly computational run for each analysis. This makes NBE and NPE ideally suited for the secure research environments, as they can provide robust Bayesian inference using minimal computational resources at the point of access.

These two methods address different inferential goals. Neural Posterior Estimators uses neural networks to directly learn an approximation of the entire posterior distribution from simulated data. In contrast, NBEs use a neural network to directly learn a point estimate (e.g. the posterior mean or median). These approaches enables efficient and flexible inference in high-dimensional or irregular settings, making it well-suited for models like MSE with missing or censored data. By training the network on synthetic datasets generated from the model, it becomes possible to recover posterior estimates even when standard analytic approaches fail.

Furthermore, this likelihood-free approach solves a fundamental estimation issue inherent in some MSE scenarios. Standard inference methods like MLEs can fail to produce a finite estimate for the population size. For example, if no individuals are observed in the overlaps between certain lists, the interaction terms in the model cannot be determined, and the likelihood function does not have a well-defined maximum (see \cite{Chan2020} for more details). In contrast, our simulation-based framework does not depend on maximizing the likelihood function. Instead, it directly learns the mapping from the data to the posterior quantities of interest (such as point estimates via NBE or the full distribution via NPE), ensuring a valid estimate can be generated even in these challenging, sparse data situations.

The amortized nature of both NBE and NPE makes it exceptionally practical for real-world applications. After an initial, one-time training cost, the trained Neural Bayes Estimator can perform inference almost instantly for any new data. This is ideal for MSE studies where data is frequently updated, as it allows for rapid re-estimation without rerunning computationally intensive algorithms like MCMC. This democratizes access to advanced Bayesian methods and removes the need for users to implement complex inference tools themselves. This is particularly valuable for researchers in secure environments like the UK's SafePods, where limited computational power and software restrictions can be a major barrier. Our tool provides a fast, efficient, and secure method for drawing insights from sensitive data.

This paper is organised as follows. We begin by describing MSE models, followed by the proposal of our Neural Estimators for MSE and how they can be applied to situations with censored data. We then evaluate the success of our method through simulation studies, before applying it to real data. Finally, we conclude with a discussion of our method and its limitations. 

\section{Multiple Systems Estimation} \label{Sec: MSE}
Consider a population of size $N$, where we are unable to observe the entire population and wish to estimate the value of $N$. Let \( K \) denote the number of lists, and assume that each individual in the population may or may not appear on each list independently. We let \( \mathcal{X} = \{0,1\}^K \setminus \{(0,0,\dots,0)\} \) represents the set of all possible non-empty capture patterns of individuals in the population across the \( K \) lists. For each capture pattern \( \bx \in \mathcal{X} \), let \( N_{\bx} \) denote the observed count of individuals with that pattern. The unobserved count \( N_0 \) corresponds to the number of individuals not captured on any list (i.e. the capture pattern $\{0, 0, \ldots, 0\})$ and is the quantity we are interested in estimating. 

A common model for MSE is the log-linear model, where we model the count $N_{\bx}$ by
\begin{equation*}
   N_{\bx} \sim \text{Poisson}(\lambda_{\bx}), \quad \text{for all } {\bx} \in \mathcal{X},  
\end{equation*}
and it is assumed that each of the counts are independent of one another. The rate parameter of this distribution is modelled through the main effects for each list, as well as the interaction between lists. It is given by
\[
\log \lambda_{\bx} = \alpha + \sum_{k \in \bx}\beta_k + \sum_{\substack{k,\, l \in \bx \\ k < l}} \gamma_{k,l} ,
\]
where \( \alpha \) is the intercept, $\boldsymbol{\beta} = (\beta_1, \ldots, \beta_K)$ is the vector of main effects for each list, and $\boldsymbol{\gamma} = (\gamma_{1,2}, \ldots, \gamma_{K-1,K})$ is the vector of interaction terms for each pair of lists. It is possible to include higher-order interactions, but in practice, interactions are usually limited to pair-wise effects due to identifiability issues and overfitting. The number of people not captured on any list is therefore model by
$$
N_0 \sim \text{Poisson}(\lambda_0),
$$
with $\log \lambda_0 = \alpha$ and $\mathbb{E}(N_0) = \exp (\alpha)$.  Under this model, the expected total population size is given by summing the observed counts and an the estimate of the unobserved count:
\[
\mathbb{E}(N) = \sum_{\bx \in \mathcal{X}} N_{\bx} + \mathbb{E}(N_0). 
\]

Although the parameter $\alpha$ is our main interest, the other parameters can be helpful in understanding how the capture process works. Positive interaction terms indicate that lists are more likely to capture the same individuals (e.g., due to shared mechanisms of detection), while negative interactions suggest avoidance or lack of overlap. 

The contribution to the likelihood function for capture pattern $\bx$ is 
$$
f_{\bx}(N_{\bx} = n_{\bx} \mid \alpha, \boldsymbol{\beta}, \boldsymbol{\gamma}) = \frac{e^{-\lambda_{\bx}}\lambda_{\bx}^{n_{\bx}}}{n_{\bx}!},
$$
with the observed data likelihood function being the product of these contributions over possible all non-empty capture patterns and the vector of observed data $\boldsymbol{n}$

\begin{equation}
    f(\boldsymbol{n} \mid \alpha, \boldsymbol{\beta}, \boldsymbol{\gamma}) = \prod_{\bx \in \mathcal{X}} f_{\bx} (N_{\bx} = n_{\bx} \mid \alpha, \boldsymbol{\beta}, \boldsymbol{\gamma}). \label{eq: MSE Poisson}
\end{equation}
Placing independent prior distributions on the model parameters, by Bayes' theorem, the posterior distribution is
$$
\pi(\alpha, \boldsymbol{\beta}, \boldsymbol{\gamma} \mid \boldsymbol{n}) \propto f(\boldsymbol{n} \mid \alpha, \boldsymbol{\beta}, \boldsymbol{\gamma})\pi(\alpha)\pi(\boldsymbol{\beta})\pi(\boldsymbol{\gamma}). 
$$

In traditional MSE applications, complete data on list overlaps are required to fit this model. However, in many real-world scenarios, some capture patterns \( \bx \) may be interval censored or missing, either randomly or due to privacy constraints, and as a consequence the likelihood function may be more computationally demanding to evaluate. This motivates the use of simulation-based approaches, as we describe in the following section.

\section{Neural Bayes Estimation for Multiple Systems Estimation Models} \label{Sec: NPE}

Instead of classical MCMC or optimization, we propose an amortized Bayesian approach. We train neural networks to approximate either specific posterior summary statistics (Neural Bayes Estimation) or the full posterior density (Neural Posterior Estimation). This involves drawing large number of samples from the prior distribution and simulating capture counts for lists based on these prior draws. We first explain how the neural methods can be applied to Multiple Systems Estimation models before introducing censored data. 

\subsection{Neural Bayes Estimators}

A Bayes estimator $\hat{\boldsymbol \theta} = \hat{\boldsymbol \theta}(\textbf{n})$, is formally defined as an estimator that minimizes the expectation of a chosen loss function, $L(\boldsymbol \theta, \hat{\boldsymbol \theta}(\textbf{n}))$, over the posterior distribution, $\pi(\boldsymbol \theta \mid \textbf{n})$. The choice of loss function determines the property of the posterior we are estimating.

For a robust point estimate, our target is the posterior median, which is the optimal Bayes estimator under the absolute error ($L_1$) loss function. It follows that the optimal estimator is given by
\begin{align}
    \hat{\boldsymbol{\theta}}^*(\textbf{n}) = \arg\min_{\hat{\boldsymbol \theta}(\textbf{n})} \mathbb{E}_{\boldsymbol{\theta} \sim \pi(\boldsymbol{\theta} \mid \boldsymbol{n})}[|\boldsymbol{\theta} - \hat{\boldsymbol \theta}(\textbf{n})|]. 
\label{eq: NBE}
\end{align}
In applications with complex likelihoods or censored count data, we cannot evaluate the posterior expectations necessary to carry out this optimisation directly. Instead, we approximate the Bayes estimator with a NBE. This is a neural network, parameterized by weights $\phi$, that is trained to approximate the mapping from the sample space to the parameter space.

Let $\hat{\boldsymbol{\theta}}_\phi(\textbf{n})$ be a neural network with weight $\phi$, in \citet{SainsburyDale2023} the authors show that the neural network $\hat{\boldsymbol{\theta}}_{\phi^*}(\textbf{n})$ approximates the Bayes estimator $\hat{\boldsymbol{\theta}}(\textbf{n})$ where the weights are defined as
\[
\phi^* \equiv \arg \min_{\phi} \frac{1}{M} \sum_{m=1}^M |\boldsymbol{\theta}^{(m)} - \hat{\boldsymbol{\theta}}_\phi(\boldsymbol{n}^{(m)})|,
\]
where $M$ is the number of simulated training samples. Each sample consists of a parameter vector $\boldsymbol{\theta}^{(m)} = (\alpha^{(m)}, \boldsymbol{\beta}^{(m)}, \boldsymbol{\gamma}^{(m)})$ drawn from the prior $\pi(\boldsymbol \theta^{(m)})$, and a corresponding simulated dataset $\textbf{n}^{(m)}$ drawn from the probability mass function in Equation (\ref{eq: MSE Poisson}).

By minimizing the $L_1$ loss over a sufficiently large and diverse set of simulated data, the network learns to approximate the posterior median for any given $\textbf{n}$. Once trained, inference is amortized: applying the estimator to a new observed dataset $\textbf{n}_\text{obs}$ requires a single, fast forward pass of the network, $\hat{\boldsymbol \theta} = \hat{\boldsymbol{\theta}}_{\phi^*}(\textbf{n}_{\text{obs}})$.

Furthermore, a key advantage of this framework is its natural extension to uncertainty quantification. By replacing the absolute error loss with the quantile loss function, $L_\tau$, we can train the neural network to target any arbitrary $\tau$-th posterior quantile. For any single parameter $\theta \in \boldsymbol \theta$, the quantile loss is defined as
$$
L_\tau(\theta, \hat{\theta}) = (\hat{\theta} - \theta)\left(\mathds{1}_{\{ \hat{\theta} > \theta \}} - \tau \right), \quad \tau \in (0,1),
$$
where $\mathds{1}$ is the standard indicator function equal to 1 if $\hat{\theta} > \theta$ and zero otherwise \citep{SainsburyDale2025}. It has been shown that estimator under the joint loss $L(\boldsymbol \theta, \hat{\boldsymbol \theta}) = \sum_{k=1}^K L_\tau(\theta_k, \hat{\theta}_k)$ is the vector of marginal posterior quantities. Hence, we are able to construct a $95\%$ credible interval by training a NBE to approximate the $\tau = 0.025$ and $\tau = 0.975$ quantiles.

\subsection{Neural Posterior Estimators}

While NBEs are trained to approximate specific posterior quantities, such as the median or quantiles, by minimizing a chosen loss function, an alternative and more comprehensive approach is NPE. The goal of NPE is not to learn a point estimate, but to approximate the entire posterior distribution $\pi(\boldsymbol \theta \mid \textbf{n})$ directly.

In this framework, we employ a conditional density estimator, $q_\phi(\boldsymbol \theta \mid \textbf{n})$, to model the posterior distribution, where $\phi$ again represents the network weights. This architecture consists of two primary components. First, the observed data \textbf{n} is passed though a multi-layer perceptron (MLP) with weights $\phi_1$ to produce a fixed size embedding vector $\textbf{s} = h_{\phi_1}(\textbf{n})$ of dimension $d^*$. This MLP acts as a summary network, designed to extract and learn informative summary statistics from the raw and potentially high dimensional count data. 

Second, a conditional normalizing flow with weights $\phi_2$ is used to model the density $q_{\phi_2}(\boldsymbol \theta \mid \textbf{s})$ between the parameters $\boldsymbol \theta$ and a simple base distribution (e.g. a standard multivariate Gaussian), where the mapping is conditioned on the summary vector $\textbf{s}$. This mapping is constructed by stacking a series of invertible affine coupling blocks. 

Similar to the NBE approach, the flow is trained on $M$ parameter and simulated data pairs $\{\boldsymbol \theta^{(m)}, \textbf{n}^{(m)} \}_{m=1}^M$ drawn from the prior distribution and the generative model. However, instead of minimizing a loss like $L_1$, the network combined weights $\phi = (\phi_1, \phi_2)$ are optimized by maximizing the log-probability of the true parameters given the simulated data. This corresponds to minimizing the negative log-likelihood loss over the training batch:
$$
\phi^* = \arg \min_{\phi} - \frac{1}{M} \sum_{m=1}^M \log q_\phi(\boldsymbol \theta^{(m)} \mid \textbf{n}^{(m)}). \label{eq: neg log lik}
$$

Once the network is trained, we provide the observed data $\textbf{n}_{\text{obs}}$ to the summary network to compute the embedding $\textbf{s}_{\text{obs}} = h_{\phi_1^*}(\textbf{n}_{\text{obs}})$. The summary vector is then used to condition the normalizing flow, yielding a fully specified approximate posterior distribution $q_{\phi^*}(\boldsymbol \theta \mid \textbf{s}_{\text{obs}})$.

From this single, trained approximation, we obtain a full characterisation of the posterior distribution. The approximate posterior distribution is typically a complex high-dimensional object and any posterior quantities of interest require integrating over the parameter space which often cannot be solved analytically. Despite this, by construction it is trivial to sample from these distributions by first sampling from the base distribution $\textbf{z} \sim p(\textbf{z})$ and applying the inverse transformation $f_{\phi^*}^{-1}(\textbf{z} \mid \textbf{s}_{\text{obs}})$. Then, by repeating this process a large number of times we may obtain samples from the approximate posterior and hence compute Monte Carlo estimates for quantities such as the mean, mean, quantiles, etc.

Another attractive feature of NPE is the ability to perform model assessment after inference through posterior predictive checks. More precisely, the posterior predictive distribution is defined as
\[
f(\textbf{n}_{\text{rep}} \mid \textbf{n}) = \int f(\textbf{n}_{\text{rep}} \mid \boldsymbol \theta) \pi(\boldsymbol \theta \mid \textbf{n}) d \boldsymbol \theta.
\]
Hence, given observed data $\textbf{n}_\text{obs}$, we may generate $S$ samples, $(\boldsymbol \theta^{(1)}, ..., \boldsymbol \theta^{(S)})$, from the approximate posterior distribution $q_{\phi^*}(\boldsymbol \theta \mid \textbf{s}_{\text{obs}}(\textbf{n}_{\text{obs}}))$ which are then used to generate new replicated datasets from the model, $\textbf{n}_{\text{rep}}^{(i)} \sim f(\textbf{n}^{(i)} \mid \boldsymbol \theta^{(i)})$ for $i=1,...,S$. 

Finally, the collection of replicated datasets $\{ \textbf{n}_{\text{rep}}^{(1)}, ..., \textbf{n}_{\text{rep}}^{(S)} \}$ are samples from the approximate posterior predictive distribution. In both cases, once trained, the neural estimator provides amortized inference and applying it to a new dataset is extremely fast since it is just a forward pass of the network. 

\subsection{Implementing the method}
For MSE data, which consists of a fixed-length non-negative vector of counts, a multi-layer perceptron (MLP) architecture is the standard choice as the data lacks the specific spatial, temporal, or permutation-invariant structures that would necessitate specialized architectures like CNNs, RNNs, or DeepSets.

For both the NBE network and the NPE encoder network, we used a fully connected architecture consisting of three hidden layers of 256 units, each utilizing ReLU activation functions. For the NBE network, the output layer activations were carefully chosen to match the posterior support for each model parameter. Specifically, real-valued parameters were assigned the Identity activation function, and parameters bounded in $[a,b]$ were assigned the shifted Sigmoid function:
\[
f(z) = a + (b-a) \cdot \frac{1}{1 + e^{-z}}.
\]
For the NPE encoder network the output layer has the identity activation and width $d^* = 128$. In order to ensure the samples generated from the approximate posterior respect the posterior support we employ a rejection sampling scheme.

We implemented both frameworks using the \texttt{NeuralEstimators.jl} package in Julia \citep{SainsburyDale2023}; Algorithm \ref{alg: Full Algo} details the full simulation and training procedure. The algorithm is general and the differences between the trained networks ultimately depend on the underlying architectures and loss function $L(\boldsymbol \theta, \hat{\boldsymbol \theta})$, which are $L_1$ and $L_{\tau}$ for the NBE approach and the negative log-likelihood. 

This training was conducted using simulation on the fly to avoid massive data storage and enhance the estimator's robustness by  simulating new parameters from the prior and data from the model during each epoch.

At each epoch, performance is assessed using a fixed validation set to calculate the validation risk. Early stopping is then employed, halting training when the validation risk plateaus or increases, thereby ensuring the final estimator generalizes effectively without overfitting. The network weights are updated using the ADAM (Adaptive Moment Estimation) optimizer, which efficiently adjusts the learning rate for each weight individually based on past gradients.

In practice, the simulated count data from the Poisson data can be very large. This is because for large population sizes, the variance of the Poisson distribution is also large. To improve the stability and accuracy of the NPE, we applied a log-transformation to the observed count data prior to training. Specifically, we used the transformation $\log(1+n_{\bx})$, which both stabilizes the variance inherent in count data and compresses large values, mitigating the influence of extreme observations. Without this adjustment, there is a considerable risk  that the estimator collapses to predicting nearly constant values across the parameter space.

Furthermore, this transformation is also important for the network's backpropagation process. By equalizing the scale of the input features, it prevents large data values from generating exploding gradients, which ensures that the weight updates during training are smooth and reliable. This stability allows the network's optimizer to converge faster and more effectively.

\begin{algorithm}
\caption{Neural Estimators for Multiple Systems Estimation}
\label{alg: Full Algo}
\begin{algorithmic}[1]
\State \textbf{Initialize:}
\begin{itemize}
    \item $M$: Number of training simulations
    \item $V$: Number of validation simulations
    \item $T$: Maximum number of epochs
    \item $\phi$: Initial network parameters
    \item $\pi(\alpha, \boldsymbol{\beta}, \boldsymbol{\gamma})$ : Prior distribution
\end{itemize}
\For{$v = 1$ to $V$}
    \State Draw $(\alpha^{(v)}, \boldsymbol{\beta}^{(v)}, \boldsymbol{\gamma}^{(v)}) \sim\pi(\alpha, \boldsymbol{\beta}, \boldsymbol{\gamma})$
    \State Generate counts $n_{\bx}^{(v)} \sim f_{\bx}(N_{\bx} \mid \alpha^{(v)}, \boldsymbol{\beta^{(v)}},\boldsymbol{\gamma}^{(v)})$ for each ${\bx} \in \mathcal{X}$ and store these as the validation set
\EndFor
\For{$t = 1$ to $T$}
    \State Draw $(\alpha^{(m)}, \boldsymbol{\beta}^{(m)}, \boldsymbol{\gamma}^{(m)}) \sim\pi(\alpha, \boldsymbol{\beta}, \boldsymbol{\gamma})$ for $m=1,...,M$
    \State Generate counts $n_{\bx}^{(m)} \sim f_{\bx}(N_{\bx} \mid \alpha^{(m)}, \boldsymbol{\beta}^{(m)},\boldsymbol{\gamma}^{(m)})$ for each ${\bx} \in \mathcal{X}$, $m=1,...,M$.
    \State Evaluate the risk $L(\boldsymbol \theta, \hat{\boldsymbol \theta})$
    \If{Early stopping criterion met}
    \State Break
    \EndIf
    \State Update network parameters $\phi$
\EndFor
\end{algorithmic}
\end{algorithm}

\subsection{Estimation with censored count data}
In the fully observed case, the data \( \mathbf{n}\) include complete counts for all non-empty capture patterns \( {\bx} \in \mathcal{X} \). For the case with censored data, we assume counts in some interval $[a,b]$ are censored  to protect individuals identities. The censored counts $n'_{\bx}$, therefore contain structural zeroes, with
$$
n'_{\bx} = \begin{cases}
n_{\bx} & \text{if } n_{\bx} \notin [a,b] \\
-1 & \text{if } n_{\bx} \in [a,b]
\end{cases}.
$$
To distinguish between structural and non-structural zeroes, we introduce a censoring (or masking) vector $\boldsymbol{c} \in \{0, 1\}^{K(K+1)/2}$, where $c_{\bx} = 1$ if the value $n'_{\bx}$ has been censored and 0 otherwise. We train the neural network on the set of censored counts $\boldsymbol{n}'$ and the censoring vector $\boldsymbol{c}$. In practice, the process is then identical to Algorithm~\ref{alg: Full Algo}, where the masking is applied after generating the data and the inputs to the neural network are now a concatenated vector $(\boldsymbol{n'_{\bx}}, \boldsymbol{c})$.

\section{Simulation Studies} \label{sec: Simulation Study}
To evidence the ability of our method to recover the model parameters and estimate the number of people not captured on any list, we carry out several simulation studies and sensitivity analyses. Throughout the rest of this paper, for models with $K$ lists we assume \textit{a priori} that
\begin{align*}
    \alpha &\sim U[1, 10] \\
    \beta_k, \gamma_{lk} & \sim N(0, 4^2),
\end{align*}
for $k=1,...,K$ and $l=1,...,k$. The priors for the $\beta$s and $\gamma$s are uninformative and the prior for $\alpha$ is chosen such that it gives hidden population sizes ranging from $e \approx 2.72$ to $e^{10} \approx 22,000$, which is considered a reasonable upper bound for MSE.

To assess the performance and robustness of our estimators, we conducted a series of simulation studies. We first established a baseline scenario for both data generation and network architecture. For the data, we assumed $K=5$ lists and a censoring interval of $[0, 10]$. The corresponding baseline NBE is a fully-connected MLP with three hidden layers, each containing 256 neurons. The baseline architecture for the NPE has a fully-connected MLP with three hidden layers, each containing 256 neurons encoder network which outputs summary statistics of dimension $d^*=128$ which is then passed through normalizing flows.

Building on this baseline, we performed a sensitivity analysis by systematically varying these parameters one at a time. We explored how performance is affected by (i) the number of lists, (ii) the censoring level, (iii) the number of hidden layers and (iv) the number of neurons in each hidden layer.

For each configuration, a new NBE and NPE was trained, and its performance was evaluated based on its ability to accurately recover the true, simulated intercept value ($\alpha$). 
In order to train each NBE, we simulated $M=10,000$ draws from the prior distributions at each epoch, generating a single dataset for each draw.

To ensure a fair assessment of generalization, a single fixed test set of new parameter-data realisations is used across all scenarios. Our primary metric is the Absolute Percentage Error (APE) for the main parameter of interest, $\alpha$, used to quantity how effectively trained estimators recover the true simulated parameter values. More precisely, let $\exp(\hat{\alpha})$ denote the inferred hidden population size, then the APE is given by
$$
\hbox{APE} = 100 \left|\frac{e^{\alpha} - e^{\hat{\alpha}}}{e^{\alpha}} \right |.
$$

We simulate $10,000$ parameters from the prior and realisations according to the model (\ref{eq: MSE Poisson}) for each of the number of lists under consideration, $K = \{3, 4, 5, 6, 10, 15\}$ and use these as the test data for each of sensitivity analyses.

Finally, to validate the statistical accuracy of our estimators, we benchmarked these estimates against those obtained from MCMC.

\subsection{Sensitivity to the population size}
\begin{figure}
    \centering
    \includegraphics[width=0.7\linewidth]{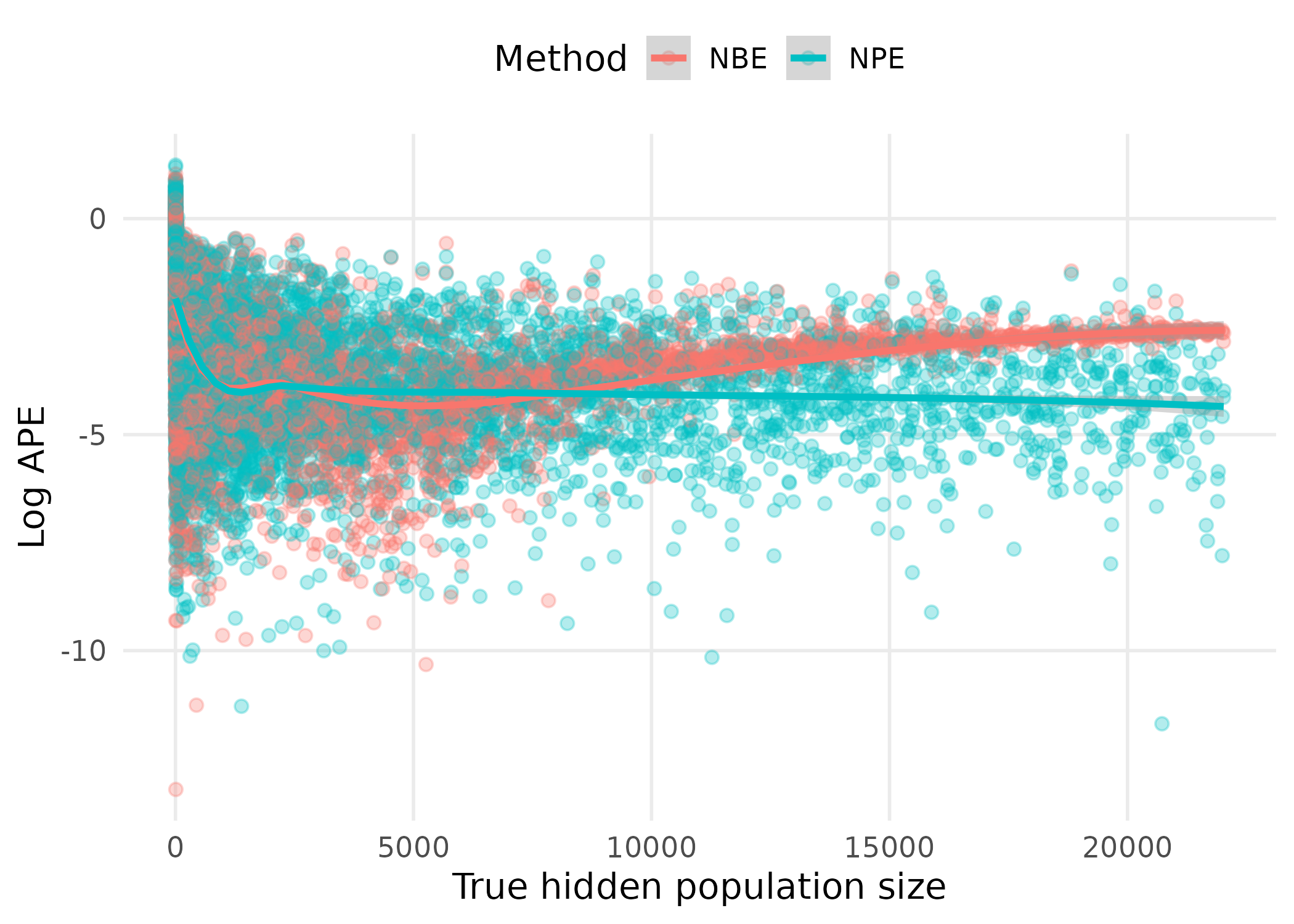}
    \caption{Log Absolute Percentage Error (Log APE) as a function of true hidden population size for Neural Bayes Estimators (NBE) and Neural Posterior Estimates (NPE). Trend lines indicate the smoothed average performance for each estimator.}
    \label{fig:population size sensitivity}
\end{figure}

We first analyze the estimators' sensitivity to the true hidden population size before varying any hyperparameters. Utilizing the estimators for $K=5$, we plot the log Absolute Percentage Error (log APE) against the true hidden population $\exp(\alpha)$ in Figure~\ref{fig:population size sensitivity}. 

For both estimators, we observe that the log APE is elevated and more variable at small population sizes $N \leq 300$. This is an expected result, as simulations with low counts generate sparse data that are insufficiently informative for precise estimation. However, once the hidden population exceeds approximately 300 individuals, the error for NPE stabilizes and remains consistent as the population size increases further.

The primary distinction between the methods emerges at a larger scale. The NBE yields similar APE than the NPE for all hidden populations up to approximately $10,000$. Beyond this threshold, the NBE's performance begins to degrade, while the NPE becomes the more accurate method. Notably, the NPE's error profile is highly stable across the entire range of population sizes tested, suggesting it is exceptionally robust to the scale of the hidden population.

The reason for this may be in part due to the NBEs ability to generalize over the training data. By construction, the data are generated with parameter sets sampled from the prior and hence $\alpha$ is sampled uniformly at random from the interval [1,10]. However, in terms of the hidden population defined by $N_0=\exp(\alpha)$, it can be shown that $N$ follows a log-uniform distribution with probability density function $f_{N_0}(n_0) = \frac{1}{9y}$ for $n_0 \in [e,e^{10}]$ and zero otherwise. As a consequence, there will be far fewer parameter sets which contain higher hidden populations and hence the estimator may struggle to effectively learn the mapping from the parameter space to sample space with less training examples.

\subsection{Varying the number of lists} \label{sec:vary_lists_sim_study}
Next, we investigate the ability to recover the model parameters under with number of lists, $K$. We investigate the performance of our method with $K = \{3, 4, 5, 6, 10\}$ lists. The first four values correspond to typical values for the number of lists in MSE studies. The largest is atypical as there are rarely situations where this number of lists exists, but also because this would result in a large number of parameters to estimate. 

\begin{figure}
    \centering
    \includegraphics[width=0.7\linewidth]{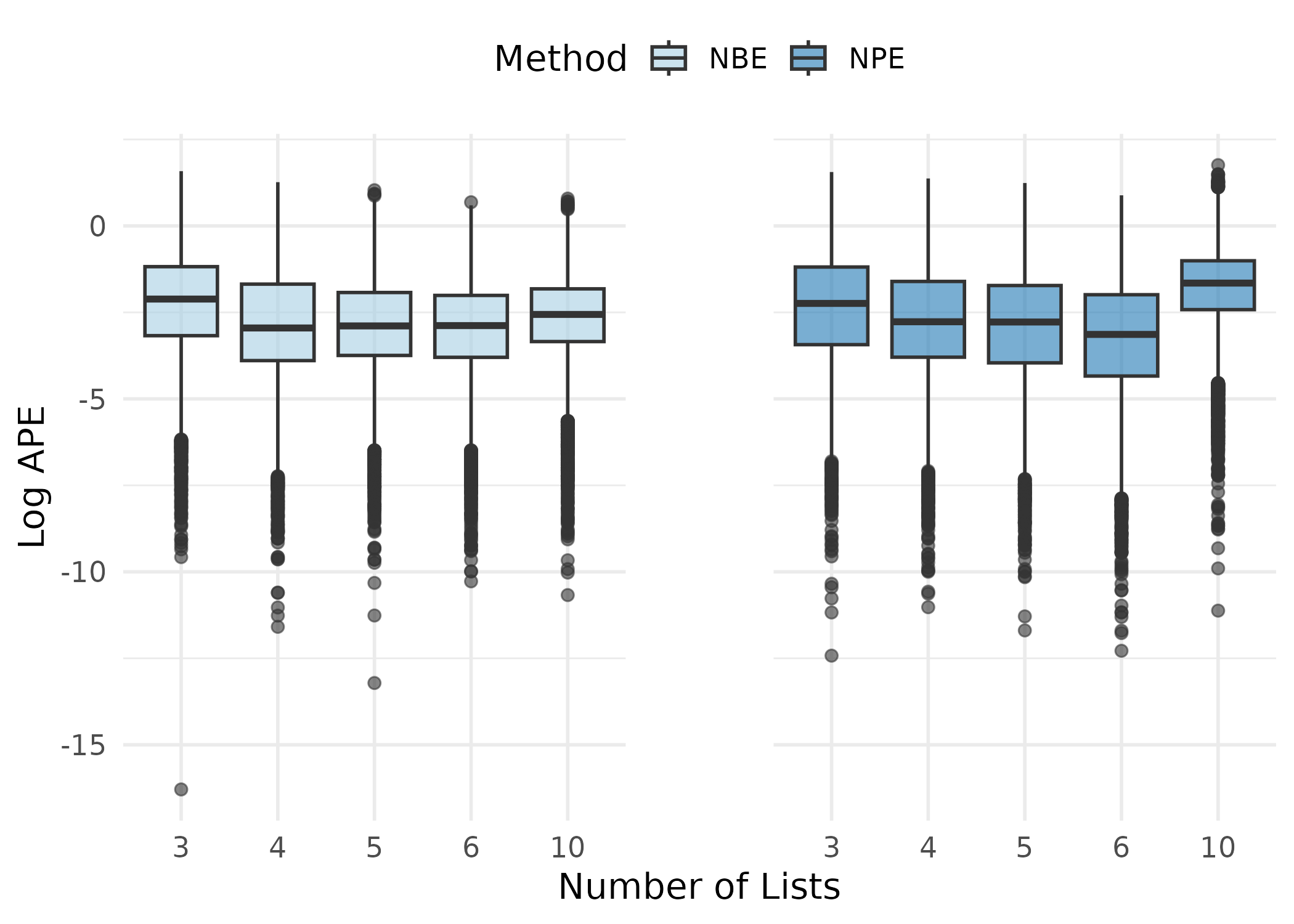}
    \caption{Comparison of NBE and NPE performance across varying list sizes. The panels display the Log APE for Neural Bayes Estimators (left) and Neural Posterior Estimates (right) calculated from the test dataset estimates (n=10,000).}
    \label{fig: sensitivity number lists}
\end{figure}

Figure \ref{fig: sensitivity number lists} shows a box-plot for the APE for each number of lists. For both estimators, the APE is similar for all lists, with a slight decrease in APE as the number of lists increases, apart for the NPE model with $K=10$ lists. Intuitively, this is due to more lists providing more data points and hence there is more information to estimate the $\alpha$ parameter which appears in each linear predictor. The exception of $K=10$ for NPE is likely due to the fact the data vector is now of dimension $\dim(\textbf{n}) = 2^{10}-1$, and hence the input to the neural network is of dimension $2 \times \dim(\textbf{n})$ since we are explicitly including the censoring vector of equal size, this may be too large of an input for the encoding network. The NPE summarises the data by the encoder network where the output vector is of dimension $d^* = 128$, which is significantly less than the inputs and therefore may not be sufficient to learn a lower-dimensional representation of the data.

\subsection{Sensitivity to the number of neurons}
We next investigate the sensitivity of the estimators to the network architecture, specifically the number of neurons per hidden layer. We conduct this analysis by training separate models with architectures using $\{8, 16, 32, 64, 128, 252\}$  neurons per layer. It is important to note that for the NPE, this parameter defines the neuron count within its encoder network.

\begin{figure}
    \centering
    \includegraphics[width=0.7\linewidth]{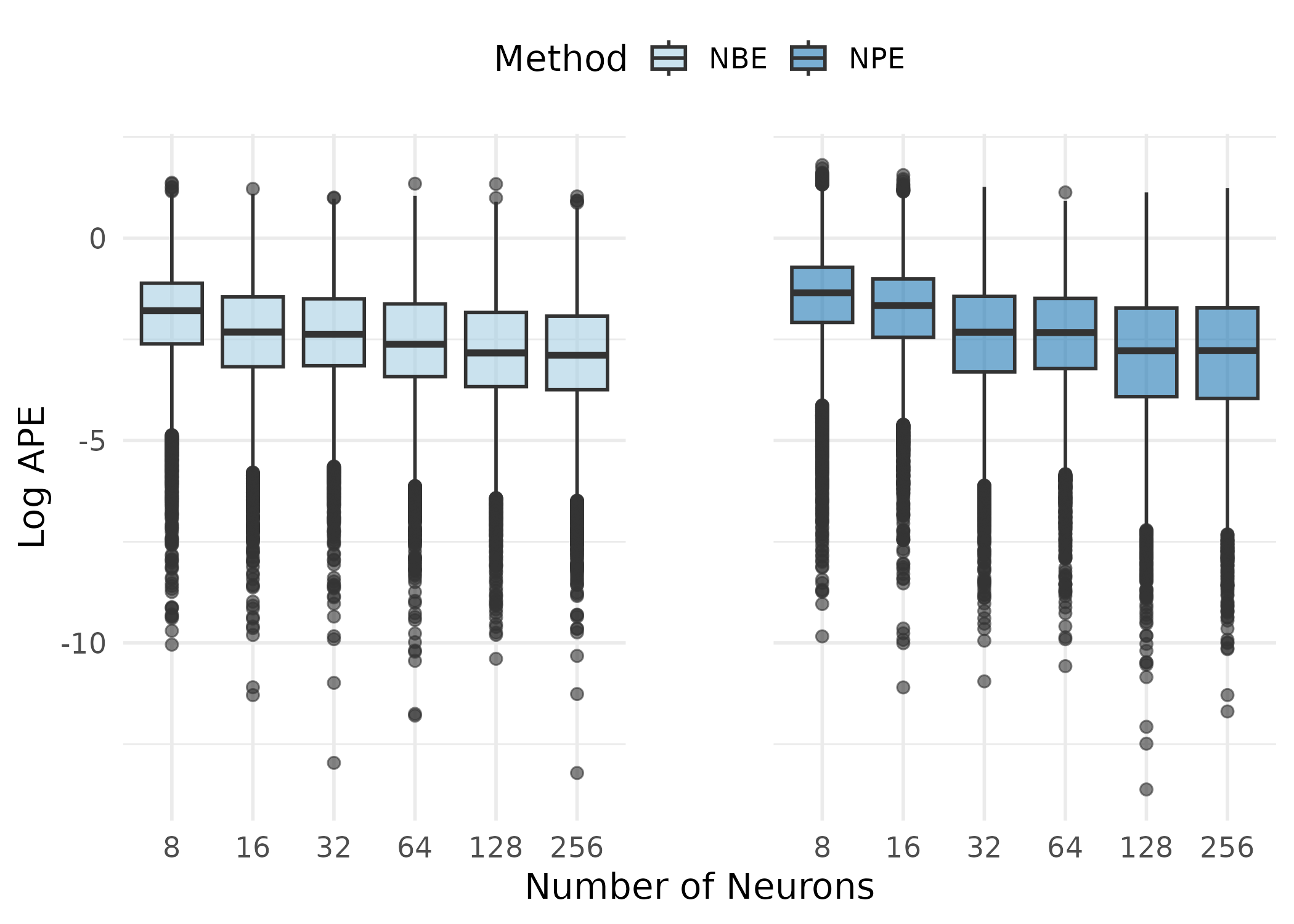}
    \caption{Comparison of NBE and NPE performance across varying number of neurons. The panels display the Log APE for Neural Bayes Estimators (left) and Neural Posterior Estimates (right) calculated from the test dataset estimates (n=10,000).}
    \label{fig: senstivity neurons}
\end{figure}

Figure~\ref{fig: senstivity neurons} displays the resulting test set Absolute Percentage Error (APE) for each configuration. A clear trend emerges for both estimators: increasing the neuron count per layer generally leads to a lower APE. This is expected, as wider layers provide a more flexible and expressive network capable of capturing complex data relationships. This finding supports our final choice of 256 neurons, which represents a practical compromise. While using more neurons (e.g., 512) might offer marginal accuracy gains, it increases computational cost and the risk of overfitting, thus yielding diminishing returns.

\subsection{Sensitivity to the number of hidden layers}
\begin{figure}[htbp]
    \centering
    \includegraphics[width=0.7\linewidth]{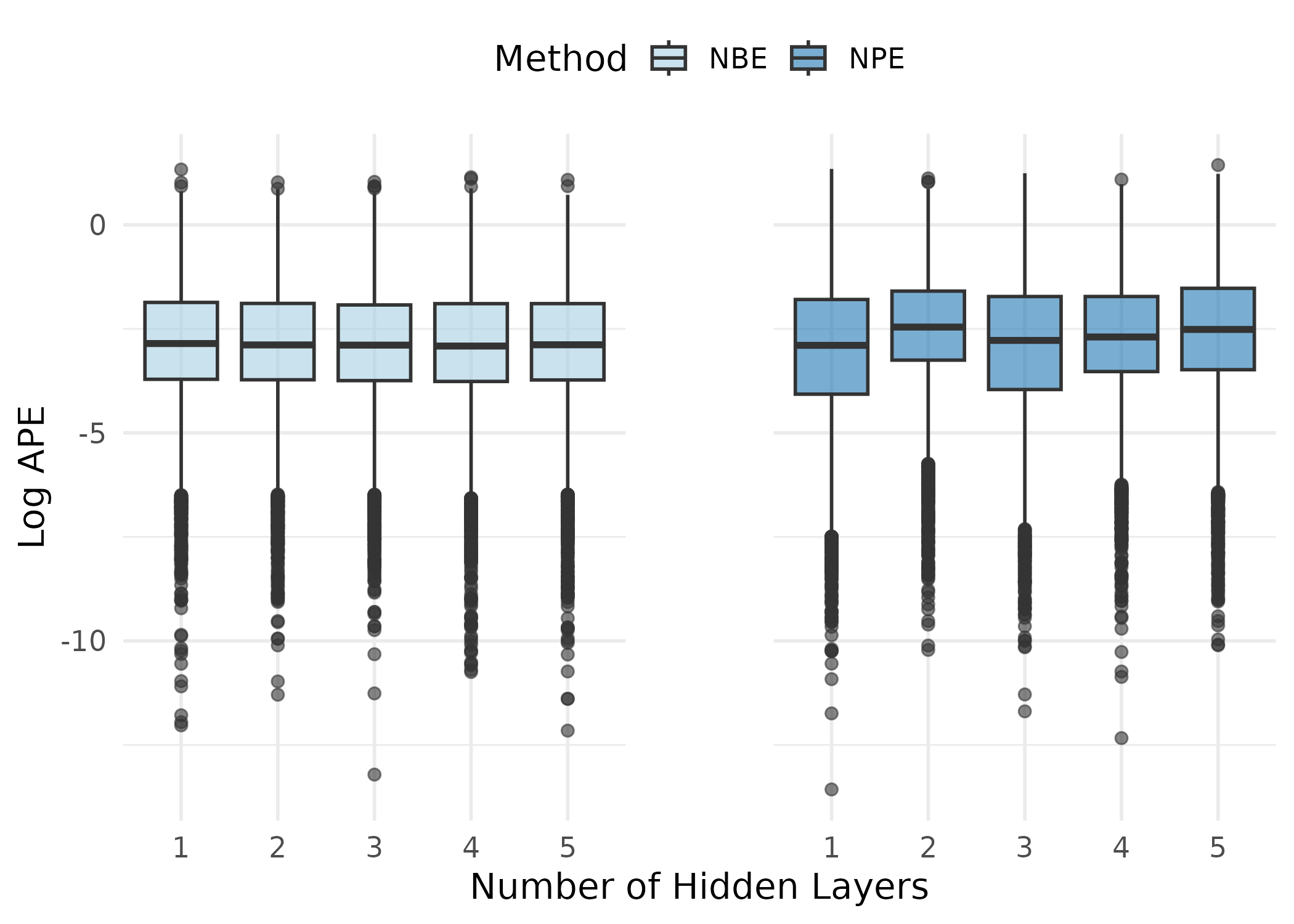}
    \caption{Comparison of NBE and NPE performance across varying number of hidden layers. The panels display the Log APE for Neural Bayes Estimators (left) and Neural Posterior Estimates (right) calculated from the test dataset estimates (n=10,000).}
    \label{fig:hidden_layers}
\end{figure}

We explored the impact of network depth, training separate estimators with $\{1, 2, 3, 4\}$ hidden layers. The results, shown in Figure~\ref{fig:hidden_layers}, indicated that performance was largely consistent across these architectures, suggesting the problem does not require an exceptionally deep model. A network with three hidden layers achieved the best performance. Since there appears to be minimal differences in the Log APE across the number of hidden layers, an architecture with three layers strikes an effective balance between model parsimony and expressive power, minimizing computational cost and the risk of overfitting

\subsection{Sensitivity to the level of censoring}
We now determine the ability of our method to recover the model parameters with different levels of censoring. For this analysis, we used the same test data previously described, but applied various censoring thresholds at $t = \{0, 2, 4, 8, 16, 32, 64, 128\}$. A separate estimator was trained for each distinct censoring level. The case of $t = 0$ corresponds to completely observed data. In practical scenarios, censoring thresholds up to 20 may be applied, but we include thresholds higher than this to identify when the method becomes unsuitable. 

\begin{figure}
    \centering
    \includegraphics[width=0.7\linewidth]{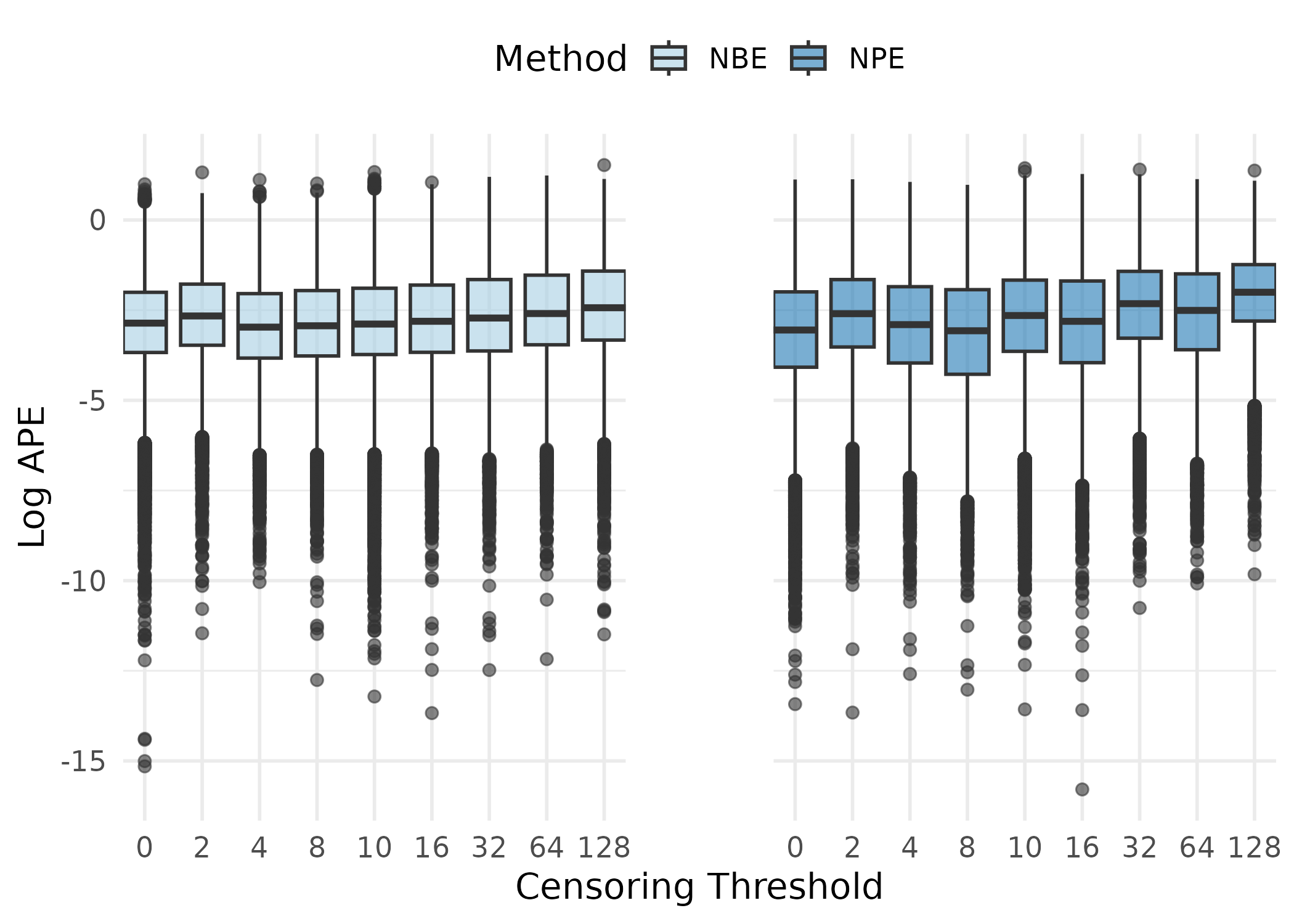}
    \caption{Comparison of NBE and NPE performance across varying upper censoring level. The panels display the Log APE for Neural Bayes Estimators (left) and Neural Posterior Estimates (right) calculated from the test dataset estimates (n=10,000).}
    \label{fig: sensitivity thresholds}
\end{figure}

Figure \ref{fig: sensitivity thresholds} illustrates the estimator's performance across various censoring thresholds. We found that no clear trend emerges, the APE remains stable regardless of the level of censoring applied to the data. While a slight increase in APE is observed at higher thresholds, this is an expected outcome as the data becomes less informative. Crucially, this effect is negligible, demonstrating the method's robustness across a wide range of data observability scenarios.

\subsection{Evaluating estimators against MCMC}

To provide a comprehensive comparison between the NBE, NPE, and posterior distribution obtained via an MCMC sampler, we evaluate various features. Recall that these methods produce outputs of different granularities:
\begin{enumerate}
    \item NBE: Produces a pre-defined set of posterior quantiles (e.g., the 2.5\%, 50\%, and 97.5\% percentiles), directly estimating the median and credible intervals.
    \item NPE: Produces samples from an approximate posterior, $q(\boldsymbol \theta \mid \textbf{n})$, from which any statistic (mean, median, quantiles) can be computed.
    \item MCMC: Produces samples from the gold-standard posterior, $\pi(\boldsymbol \theta \mid \textbf{n})$, which also allows any statistic to be computed.
\end{enumerate}

Our evaluation framework therefore aims assesses three aspects of performance: (1) point estimate accuracy, (2) quality of uncertainty quantification, and (3) computational cost. In this simulation study, we have two ``ground truths'': the simulated true hidden population ($N_0^{\text{true}} = e^\alpha$), which serves as the benchmark for accuracy, and the MCMC posterior $\pi(\boldsymbol \theta \mid \textbf{n})$, which serves as the gold-standard benchmark for uncertainty, assuming that the chains have converged.

We consider the same test data previously described from $K=5$ lists with a censoring interval of $[0,10]$ and compare the methods using the criteria described above. Similar to \cite{King21}, let $\mathcal{Y}$ denote the set of capture patterns with censored observations, i.e. the true observed count lies somewhere on the interval $[0,10]$. Under this scenario the probability mass function of the data is therefore
$$
f(\boldsymbol{n} \mid \alpha, \boldsymbol{\beta}, \boldsymbol{\gamma}) = \prod_{\bx \in \mathcal{X}} f_{\bx}(N_{\bx} = n_{\bx} \mid \alpha, \boldsymbol{\beta}, \boldsymbol{\gamma}) \prod_{\bx \in \mathcal{Y}} F_{\bx}(N_{\bx} = n_{\bx} \mid \alpha, \boldsymbol{\beta}, \boldsymbol{\gamma}),
$$
where $F_{\bx}$ denotes the cumulative distribution function. Note that in general for any censored interval $[a,b]$ where $0 < a < b$, the contribution of the censored capture patterns becomes $\mathbb{P}(a \leq N_{\bx} \leq b)$ for all $\bx \in \mathcal{Y}$.

The MCMC analysis was implemented in Julia using the \texttt{Turing.jl}, a flexible and powerful package for probabilistic programming \citep{TuringJulia}. We employed the No-U-Turn Sampler (NUTS) \citep{hoffman2011_NUTS}, an efficient Hamiltonian Monte Carlo algorithm, to perform the sampling.

For each of the 10,000 simulated datasets, we ran four independent MCMC chains for 5,000 iterations each, discarding the first 1,000 iterations as burn-in. We assessed convergence using the Gelman-Rubin statistic $(\hat{R})$ \citep{Gelman1992}. We considered chains to have successfully converged only if $\hat{R} \leq 1.01$ for all model parameters.

\subsection{Point estimate accuracy}

To make a fair comparison of all three methods, we first select a single point estimate for each. We use the posterior median, as it is a robust estimate for potentially skewed posteriors and is directly output by the NBE.

Our primary focus is the ability for the model to recover the true hidden population size, therefore for each of the test datasets we compare our estimates, $\hat{N}_0^{\text{NBE}}$, $\hat{N}_0^{\text{NPE}}$ and $\hat{N}_0^{\text{MCMC}}$ with the truth $N_0^{\text{true}}$. For each estimate $\hat{N}_0$, we quantify the performance using raw error, defined as $\hat{N} - N_0^{\text{true}}$, the absolute error (AE), $|\hat{N} - N_0^{\text{true}}|$, and the APE.

The raw error distribution is used to assess estimator bias. The Absolute Error quantifies the magnitude of the estimation error. The APE is arguably the most critical metric for this study, as it normalizes the error by the true population size, permitting a meaningful comparison of relative accuracy across simulations with different scales.

Analysis of the raw error (left plot) confirms that all three estimators are median-unbiased, with distributions centered tightly at zero. All methods are subject to extreme outliers, though the NBE appears the most robust with smaller tails. The NPE, in contrast, exhibits a slight positive median bias (see Supplementary Material, Figure 1 for full-range plots).

The Absolute Error (center plot) distributions appear similar for all three methods, with the majority of errors concentrated near zero. The NBE maintains a marginal advantage, with its distribution being slightly more concentrated at low error values.

The APE (right plot) reveals the most significant performance distinctions. The MCMC distribution is visibly wider and has a much larger tail. Conversely, both neural methods demonstrate superior performance with their APE distributions concentrated near zero, reflecting a lower median APE and substantially smaller variance. This demonstrates that the neural estimators are more robust and provide more accurate relative estimates across the range of population scales tested. Finally, these metrics are summarised by Bias, Mean Absolute Percentage Error (MAPE) and Root Mean Squared Error (RMSE) in Table~\ref{tabel:comparison_of_methods}.

\begin{figure}[htbp]
    \centering
    \includegraphics[width=0.8\linewidth]{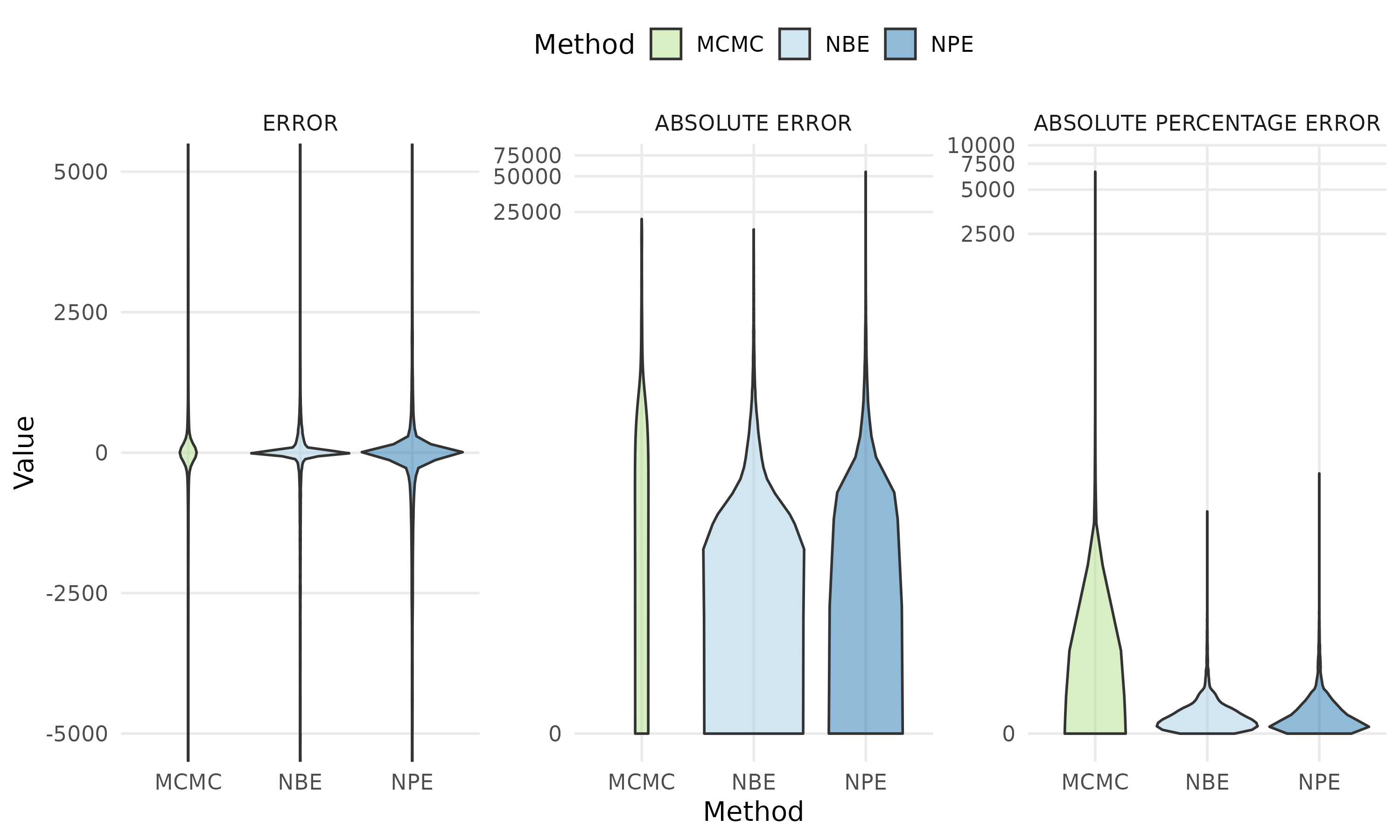}
    \caption{Violin plots showing the distributions of per-simulation error metrics for the estimated hidden population ($\hat{N}_0$). Each plot compares the MCMC (green), NBE (light blue), and NPE (dark blue) parameter estimates. The metrics are: (left) raw error, (middle) absolute error and (right) absolute percentage error. Note that for the error plot the y-axis has been restricted for clarity and the absolute error and absolute percentage error are displayed on the log scale.}
    \label{fig:point_estimate_comparison_zoomed}
\end{figure}

\subsection{Uncertainty quantification}

We next assess the quality of the uncertainty quantification  for each method. This evaluation is split into two parts: a direct comparison of 95\% credible interval (CI) coverage, which all three methods can produce, and a more detailed analysis of full posterior calibration, which is only possible for the NPE and MCMC methods.

The primary metric for uncertainty quantification accuracy is calibration. An estimator $\hat{\theta}$ is well-calibrated if its 95\% credible intervals contain the true parameter $\theta$ in 95\% of repeated simulations \citep{LittleCalibration}. We first compute the Empirical Coverage Probability (ECP) for the 95\% CI from all three methods (Table~\ref{tabel:comparison_of_methods}). For the NBE the 95\% is taken directly from the network's outputs. For MCMC and NPE we generate samples from the posterior and then compute the quantiles.

\begin{figure}[htbp]
    \centering
    \includegraphics[width=0.6\linewidth]{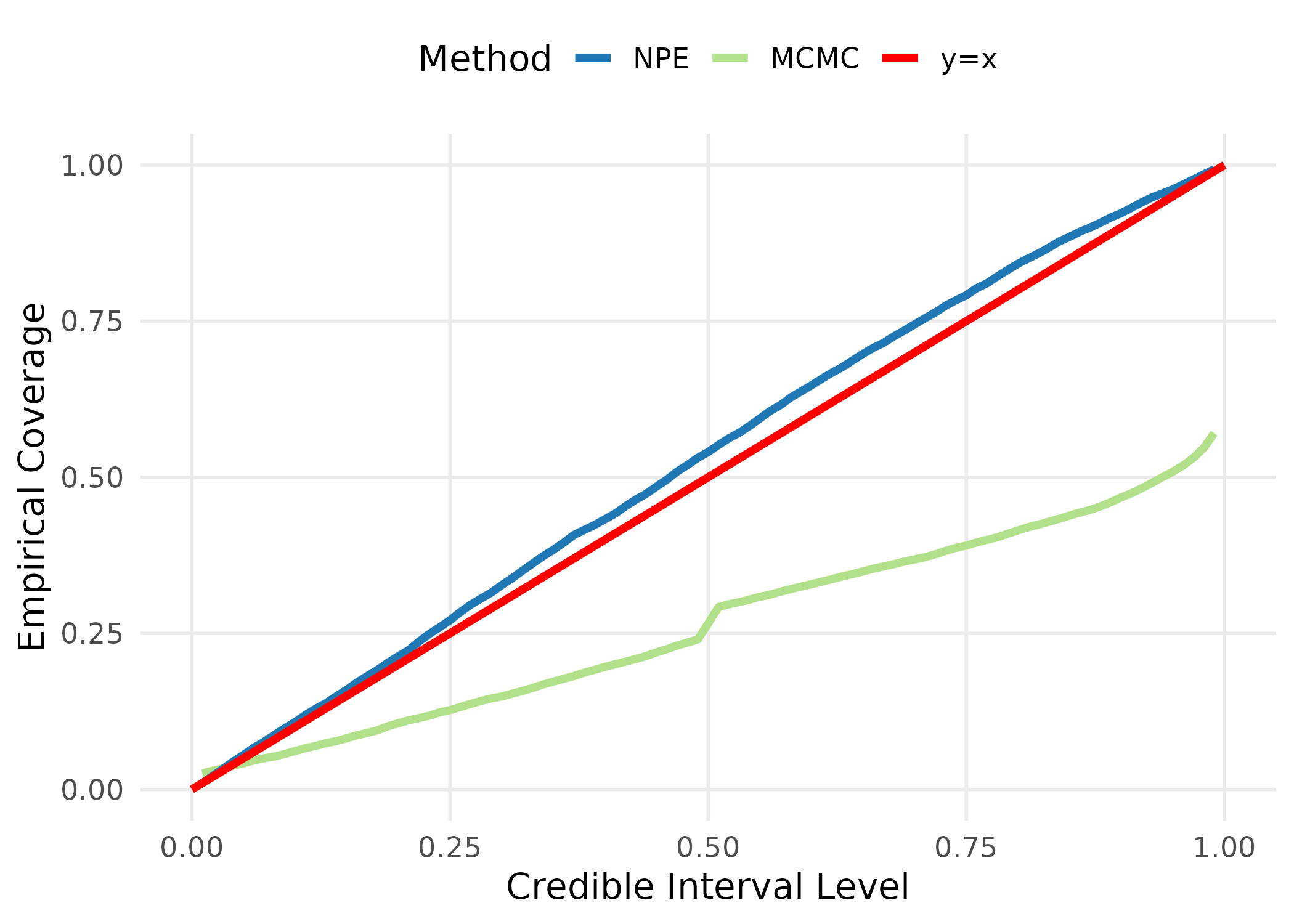}
    \caption{Calibration plot comparing the empirical coverage (y-axis) of NPE (blue) and MCMC (green) against the nominal credible interval level (x-axis). The red y=x line indicates perfect calibration.}
    \label{fig:coverage_comparison}
\end{figure}

Additionally, since NPE and MCMC produce full posterior samples, we can assess their calibration across the entire range of credible intervals, not just at 95\%. We generate full calibration plots by computing the ECP for regular intervals (Figure~\ref{fig:coverage_comparison}). The plot indicates that the NPE is well calibrated, indicated by closely following the $y=x$ which suggests the posterior distributions are honest with regards to uncertainty. At the 95\% level, NPE achieves an empirical coverage of 0.961, which is close to ideal and slightly conservative.

On the other hand, the MCMC exhibits significant miscalibration. The empirical coverage curve lies significantly below the $y=x$ line, demonstrating that the posteriors are in general too narrow and overconfident. Furthermore, at the 95\% level the MCMC credible intervals contained the true parameter 50.9\% of the time, suggesting that the sampler is failing to explore the full posterior.

Finally, the NBE's 95\% credible intervals are slightly larger than the ECP for NPE, suggesting that they both provide reasonable accurate and reliably calibrated uncertainty.

\subsection{Computational speed}

A key practical advantage of the neural methods is their amortized computational cost. This cost is split into two phases: a one-time `training' cost and a `per-inference' cost.

\begin{figure}[htbp]
    \centering
    \includegraphics[width=0.6\linewidth]{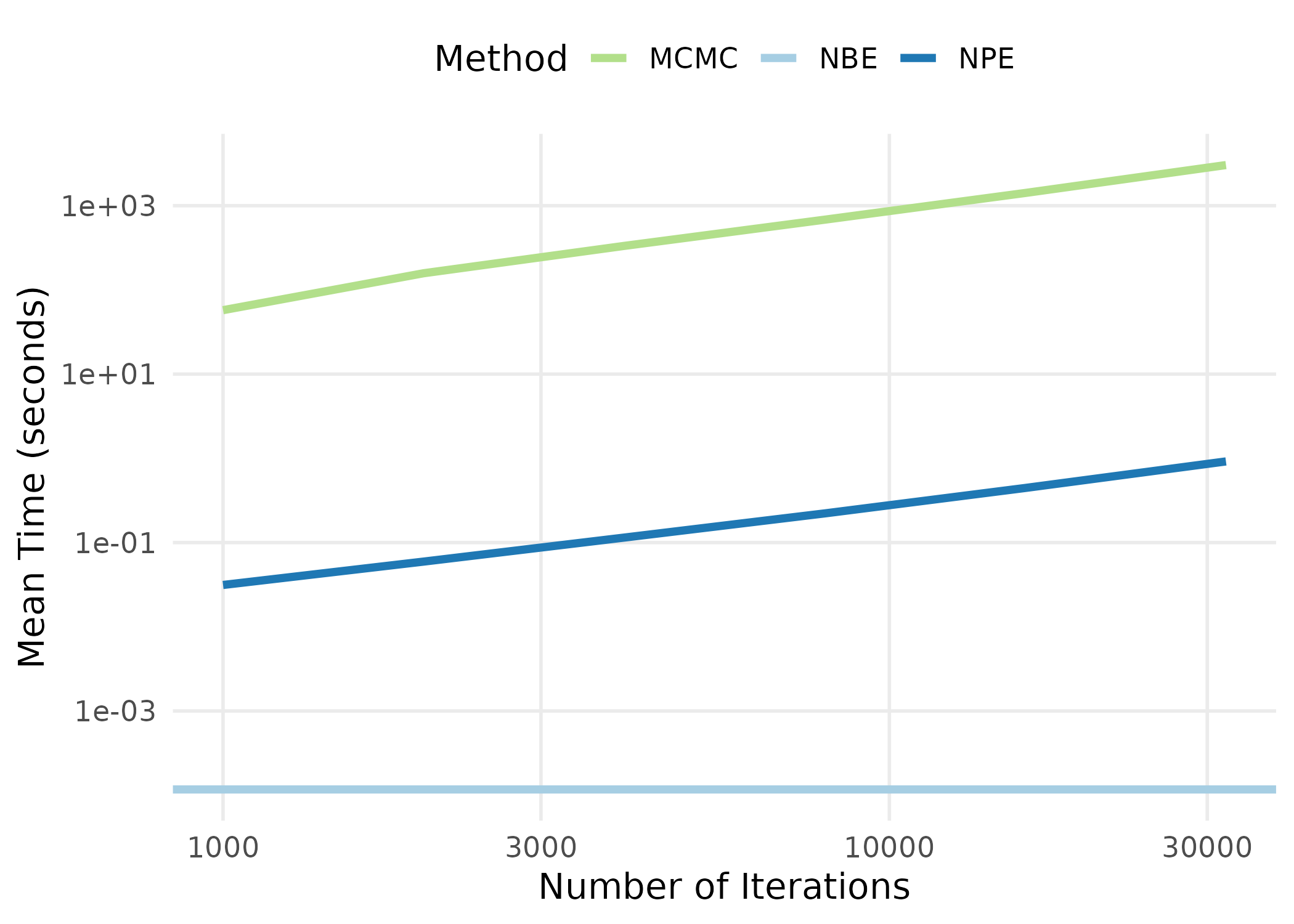}
    \caption{Mean inference time (seconds, log scale) as a function of the number of iterations for MCMC (green), NBE (light blue), and NPE (dark blue). Each point represents the average inference time over 100 test datasets, with each run repeated 10 times. The plot highlights the substantial efficiency gains of amortized inference, with NBE showing near-constant, minimal time regardless of iterations, and NPE exhibiting significantly faster inference than MCMC.}
    \label{fig:speed_comparison}
\end{figure}

The MCMC sampler has no training cost, but its inference cost is high, requiring minutes to hours to converge for each new dataset. As Figure~\ref{fig:speed_comparison} shows, MCMC's runtime (green) scales directly with the number of iterations, making it computationally expensive.

The NBE and NPE, conversely, require a significant, one-time training investment. Once trained, the cost of inference for a new dataset is straightforward, requiring only forward passes of the neural network network.

The NBE (light blue) and NPE (dark blue) inference times are several orders of magnitude faster than MCMC's. This amortized efficiency makes them far superior for large-scale simulation studies or real-time applications.

By fitting a linear model to the data, we can observe the scaling properties. Both MCMC and NPE exhibit a cost that scales linearly with the number of iterations, $N$, i.e., $\mathcal{O}(N)$. In contrast, the NBE is an $\mathcal{O}(1)$ operation, with a runtime independent of $N$.

The MCMC sampler (green) is the most expensive, with a cost of approximately 46.8 ms per iteration. The NPE (dark blue) is dramatically more efficient, costing only 0.032 ms per iteration. The NBE (light blue) is in general the fastest, requiring a constant 0.118 ms for a single forward pass. This means that at inference time, the NPE is on average approximately 1,450 times faster than the MCMC sampler, highlighting the practical advantage of amortized neural estimators.

\begin{table}
\centering
\begin{tabular}{lrrr}
\toprule
  & NBE    & NPE  & MCMC  \\
\midrule
Bias & -636 & -194 & 2061 \\
MAPE & 0.53 & 0.669 & 52.1  \\
RMSE & 2190 & 2032 & 6593 \\
95\% ECP & 0.966 & 0.960 & 0.493 \\
Training time in s & 1051 & 1577 & 0 \\
Inference time $(N)$ in ms & 0.117 & $0.0285 \times N$  & $93 \times N$ \\
\bottomrule
\end{tabular}
\caption{Summary of average point estimation error, empirical coverage probability and computational costs (in ms) for each method. }
    \label{tabel:comparison_of_methods}
\end{table}

\subsection{MCMC convergence}

\begin{figure}
    \centering
    \includegraphics[width=0.7\linewidth]{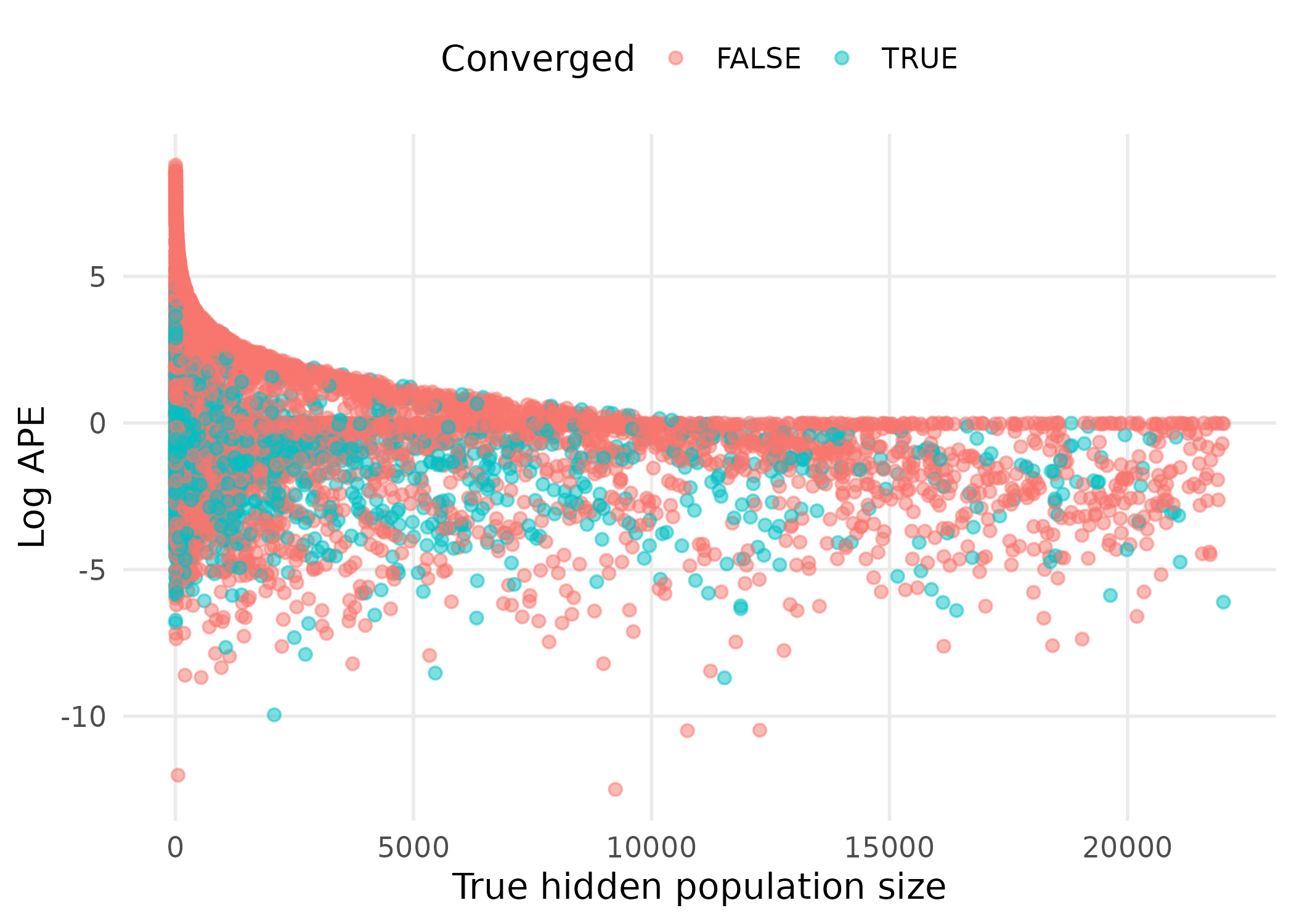}
    \caption{A scatter plot showing the true hidden population size against the log APE computed from posterior samples obtained via MCMC where each sample is (blue) converged or (red) not converged, where convergence is defined as the Gelman-Rubin (GR) statistic $\leq 1.01$.}
    \label{fig:mcmc_ape_convergence}
\end{figure}

An important finding in the simulation study was the MCMC's high failure rate. Only 7,141 of the 10,000 (71.41\%) test datasets achieved convergence and it can be seen (Figure~\ref{fig:mcmc_ape_convergence}) that posteriors that failed to converge to a stationary distribution exhibited significantly larger absolute percentage errors.

\begin{figure}
    \centering
    \includegraphics[width=0.7\linewidth]{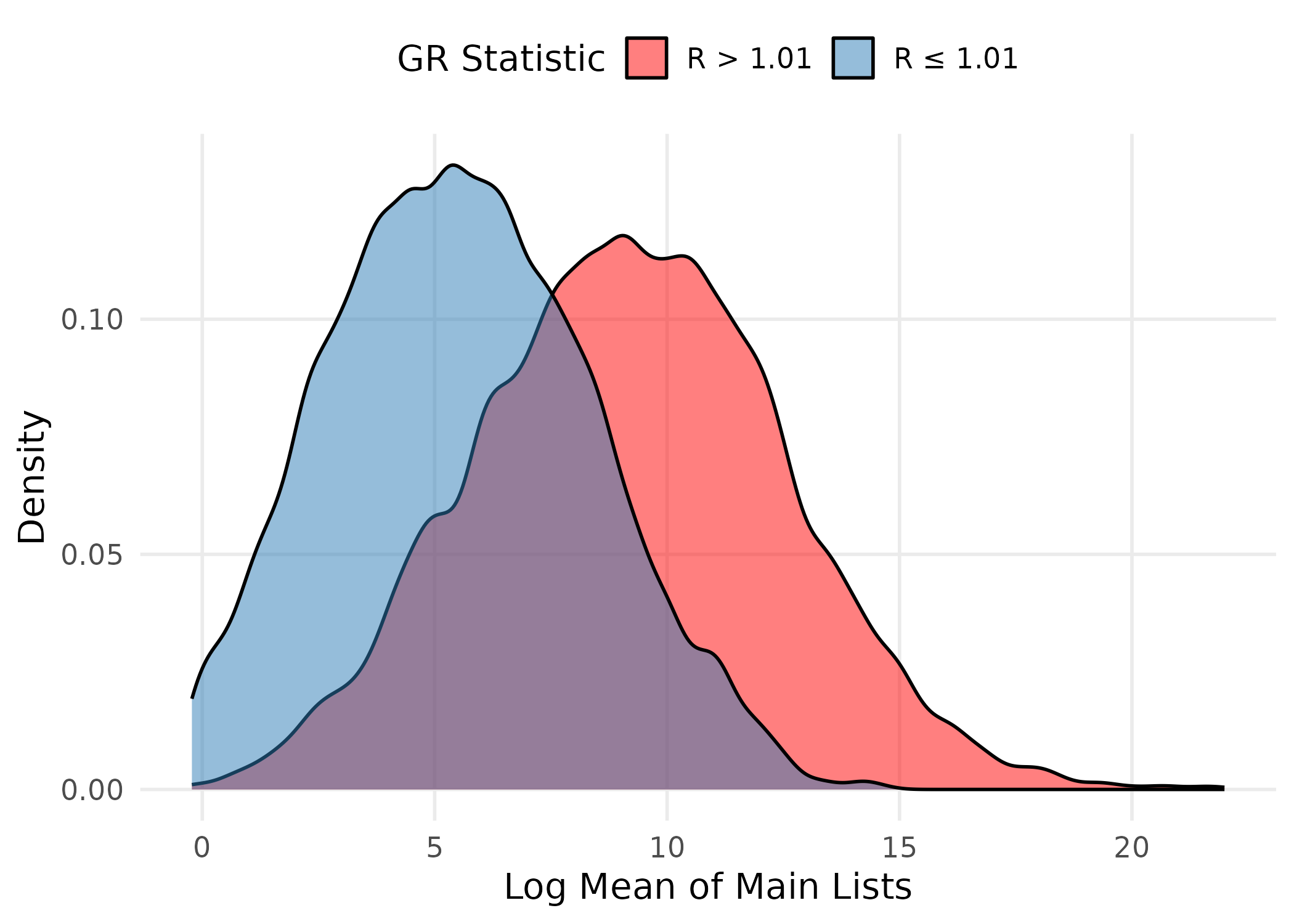}
    \caption{Densities of the log average count on each of the five main lists for chains that have converged (blue) overlaid with those not converged (red), where convergence is defined as the Gelman-Rubin (GR) statistic $\leq 1.01$.}
    \label{fig:nbe_mcmc_convergence}
\end{figure}

Upon inspection, MCMC failures were concentrated in atypical datasets, particularly those simulations generating unusually large list counts (Figure~\ref{fig:nbe_mcmc_convergence}). 

These data likely induce complex posterior geometries that proved difficult for the NUTS sampler which often manifested as an inability for the sampler to find correct tuning parameters during the warm-up phase. While it might be possible to remedy these failures with expert implementation and careful, case-by-case tuning, this highlights that the MCMC solution does not work `out of the box' for all data.

\begin{figure}
    \centering
    \includegraphics[width=\linewidth]{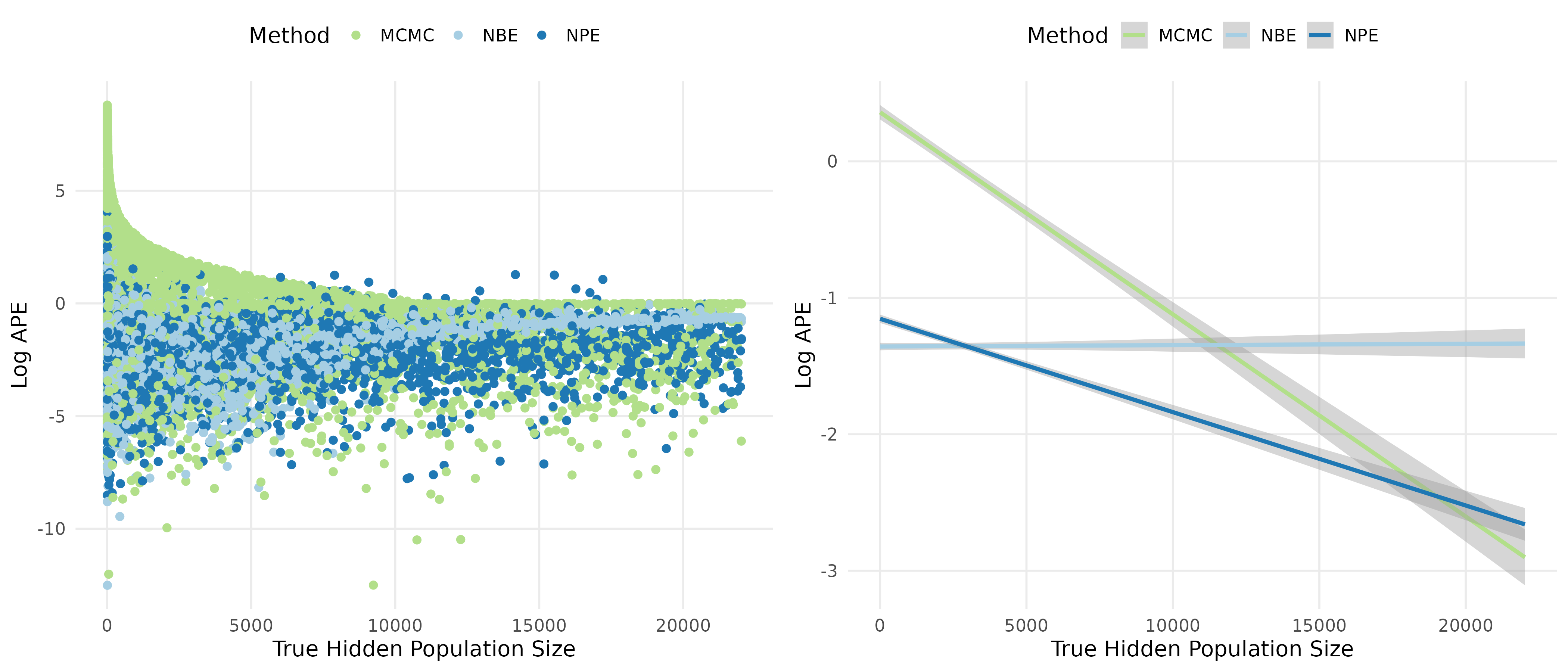}
    \caption{Left: The log absolute percentage error of the inferred hidden population size against the true population size for NBE (light blue), NPE (dark blue) and MCMC (green). Right: A linear fit for each method with 95\% confidence intervals.}
    \label{fig:nbe_mcmc_intercept}
\end{figure}

For the 7,141 converged datasets, we compared the posterior median of the intercept parameter from MCMC with the neural methods estimates (Figure~\ref{fig:nbe_mcmc_intercept}). This comparison validates the NBE's accuracy in reproducing the results of the more computationally intensive MCMC.

We evaluate the performance using APE and find that for small population sizes, the NBE achieved a lower average APE than MCMC. However, for very large population sizes, MCMC performs better, likely benefiting from more informative data. It can also be seen that NBE's APE remained reasonably consistent across all population sizes, demonstrating robust and stable performance irrespective of the true hidden population size.

The MCMC's significant convergence failures, even in simulations that may be rare in practice, highlights the robustness of the estimator. The amortized estimator provides reliable estimates across a wide range of data-generating parameters, including challenging scenarios where traditional samplers fail.

\section{Real Data Analysis} \label{sec: Real Data}
We now use our method on two real datasets and estimate the size of the hidden population. 

\subsection{Modern Slavery in the UK}
We apply our method to data on the number of victims of modern slavery in the UK, which is analysed in \citet{Silverman2020}. This dataset comprises victim counts recorded by distinct reporting mechanisms, including Local Authorities, Non-Governmental Organisations, and Police Forces. This is a notable example of an MSE dataset, as the original study findings contributed to the drafting of new legislation in the UK.

We trained various architectures for a system of $K=5$ lists with no censoring. The final models were selected by minimizing the Mean Absolute Percentage Error (MAPE) of $\alpha$ on the unseen test data. The selected NBE model utilizes four hidden layers of 256 neurons each, while the optimal NPE model employs an encoder network with one hidden layer of width 128, where the summary network output also has a dimension of 128.

\begin{figure}
    \centering
    \includegraphics[width=0.7\linewidth]{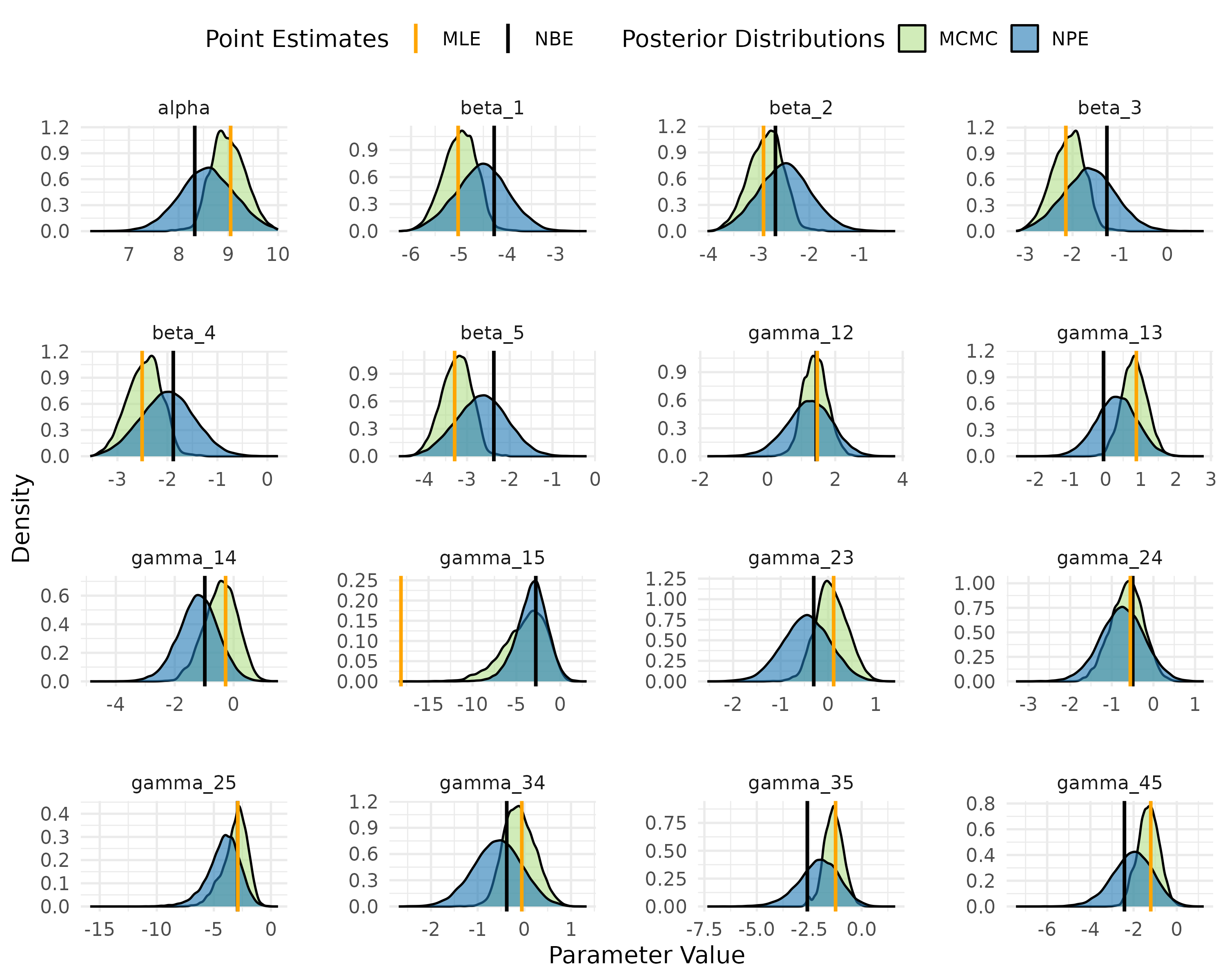}
    \caption{Comparison of parameter estimates and posterior distributions for the Modern Slavery dataset. Vertical lines represent point estimates (MLE and NBE), while filled curves show the posterior densities (MCMC and NPE).}
    \label{fig:silverman_par_estimates}
\end{figure}

A comparison of the inferred posterior distributions and point estimates is shown in Figure~\ref{fig:silverman_par_estimates}. For completeness, we also compare estimates inferred through neural methods against MCMC and MLE. Our model estimates are broadly consistent with those reported in \citet{Silverman2020}; however, we utilize all pairwise interaction terms without performing formal model selection. Notably, the MLE for the interaction term $\gamma_{15}$ is significantly more negative than the Bayesian estimates. This occurs because there is zero overlap between lists 1 and 5, rendering the parameter unidentifiable in a likelihood framework. The Bayesian and neural methods handle this more robustly due to regularization from the prior.

Table~\ref{table:silverman_hidden_pops} compares the inferred hidden population estimates $\hat{N}_0 = \exp(\hat{\alpha})$ across two estimates using the likelihood function (MLE and the posterior mean) and both of our neural estimators. For the Bayesian methods, we report the posterior median and the 95\% credible interval, for the frequentist approach, we report the maximum likelihood estimate and the 95\% confidence interval.

We find that all estimates are broadly similar, however the methods diverge significantly in their estimation of the intercept $\alpha$. As seen in Figure~\ref{fig:silverman_par_estimates}, the MCMC posterior for $\alpha$ is centered approximately at 9.0, aligning closely with the MLE, whereas the NPE and NBE distributions are shifted left, centering closer to 8.5. Due to the exponential link function $\hat{N}_0 = \exp(\hat{\alpha})$, this modest shift in log-space translates to a large difference in population scale (approx. 8,000 vs 5,000).

\begin{table}
\centering

\begin{tabular}{lcc}
\hline
Method & Hidden Population & 95\% CI \\
\hline
MCMC & 7987 & (4462, 16154) \\
NPE & 5291 & (1784, 15294) \\
NBE & 4123 & (889, 18746) \\
MLE & 8489 & (3761, 19159) \\
\hline
\end{tabular}
\caption{Inferred hidden populations and uncertainty for each of the methods for the modern slavery dataset.}
\label{table:silverman_hidden_pops}
\end{table}

\begin{figure}
    \centering
    \includegraphics[width=\linewidth]{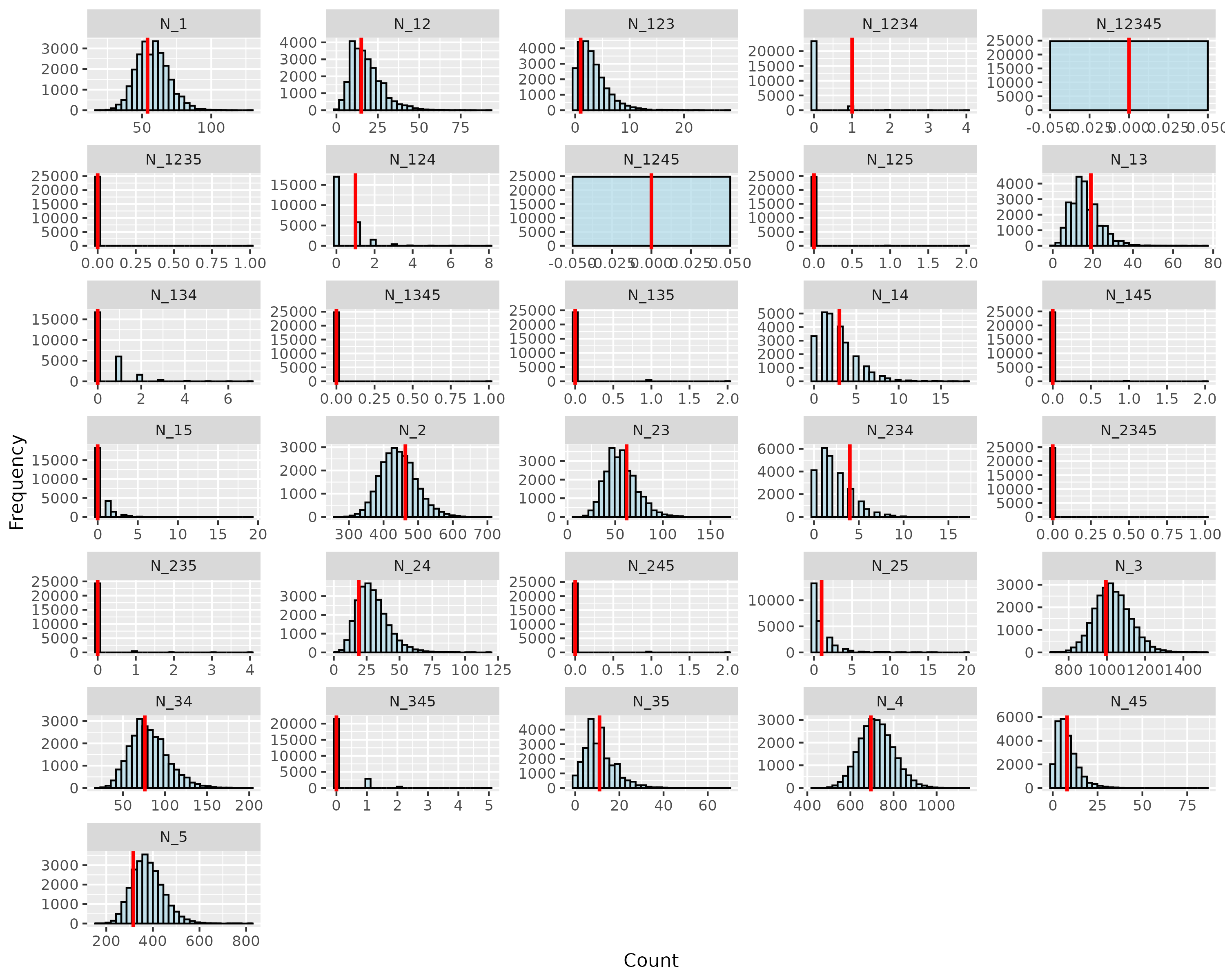}
    \caption{Posterior Predictive Distributions (PPD) generated using NPE. Histograms show the simulated counts for different list intersections, while the vertical red lines indicate the actual observed counts in the dataset.}
    \label{fig:silverman_ppd}
\end{figure}

Finally, a key advantage of the Bayesian paradigm is the ability to assess model fit via the posterior predictive distribution (PPD), $\pi(\textbf{n}_{\text{new}} \mid \textbf{n})$, which represents the distribution of new data implied by the model conditional on the observed data. We generated PPDs by simulating new datasets using posterior samples obtained from the NPE. As shown in Figure~\ref{fig:silverman_ppd}, the observed data counts (indicated by the red lines) fall within the 95\% credible intervals of the marginal PPDs for nearly all intersection profiles. The notable exception is the count for the four-list overlap $n_{1234}$. Aside from this outlier, the model demonstrates a reasonable fit to the observed data.

\subsection{Female Drug Users in North East England}
The second application concerns data on female drug users in North East England, originally published by \citet{King21}. The dataset records the number of female drug users aged 15-34 identified by four distinct reporting mechanisms: Probation, Drug Intervention Programme (DIP) prison assessments, Drug treatment, and DIP community assessments.

While the number of women appearing on each list and various list combinations is available, counts between 1 and 4 were censored to prevent the potential disclosure of identities. We trained various architectures for this system of $K=4$ lists, incorporating the censoring interval $[1,4]$ into the networks inputs. As in the previous section, the final models were selected by minimizing the Mean Absolute Percentage Error (MAPE) on unseen test data. The optimal NBE model utilizes four hidden layers of 256 neurons each, while the selected NPE model employs an encoder network with one hidden layer of width 128, with a summary network output dimension of 128.

\begin{figure}
    \centering
    \includegraphics[width=\linewidth]{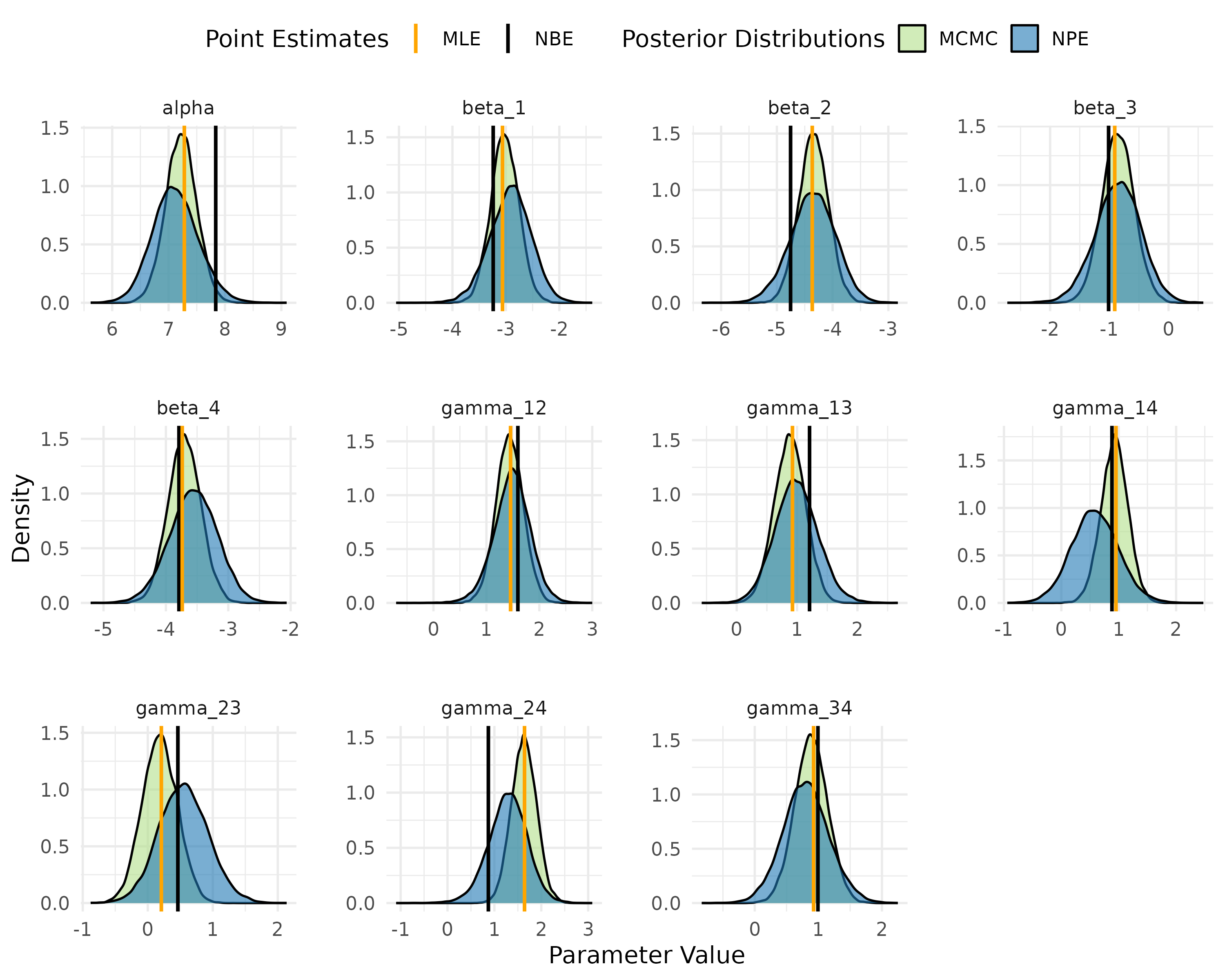}
    \caption{Comparison of parameter estimates for the Female Drug Users dataset ($K=4$). The vertical lines indicate point estimates (MLE and NBE), while the filled curves show posterior densities (MCMC and NPE).}
    \label{fig:king_par_estimates}
\end{figure}

Parameter estimates and posterior distributions are displayed in Figure~\ref{fig:king_par_estimates}. Unlike the Modern Slavery dataset, where the NBE produced a conservative population estimate, here the NBE point estimate for the intercept $\alpha$ (black vertical line) is noticeably higher than the Maximum Likelihood Estimate (orange vertical line). This shift suggests that the neural Bayes estimator infers a larger hidden population. This is consistent with the observation that the NBE estimates for the main effects (e.g., $\beta_1, \beta_2$, etc) are generally more negative than their MLE counterparts, implying lower capture probabilities and, consequently, a larger unobserved population.

Additionally, we observe that the NPE posterior distribution (blue density) for the interaction terms $\gamma_{14}, \gamma_{23}$ and $\gamma_{24}$ diverges from the MCMC posterior (green density). This suggests that the neural posterior estimator detects difference dependencies between the lists than the standard MCMC approach.

Table~\ref{table:king_hidden_pops} compares the inferred hidden populations. Consistent with the parameter estimates, the NBE produces a higher estimate than the classical methods. Note that for the censored data, the likelihood and posterior evaluations explicitly account for the probability mass assigned to the interval $[1,4]$.

\begin{table}
\centering
\begin{tabular}{lcc}
\hline
Method & Hidden Population & 95\% CI \\
\hline
MCMC & 1371 & (795, 2346) \\
NPE & 1225 & (579, 2852) \\
NBE & 2530 & (80, 3942) \\
MLE & 1450 & (840, 2505) \\
\hline
\end{tabular}
\caption{Inferred hidden populations ($\hat{N}_0$) and uncertainty intervals for the \citet{King21} dataset.}
\label{table:king_hidden_pops}
\end{table}

\begin{figure}
    \centering
    \includegraphics[width=\linewidth]{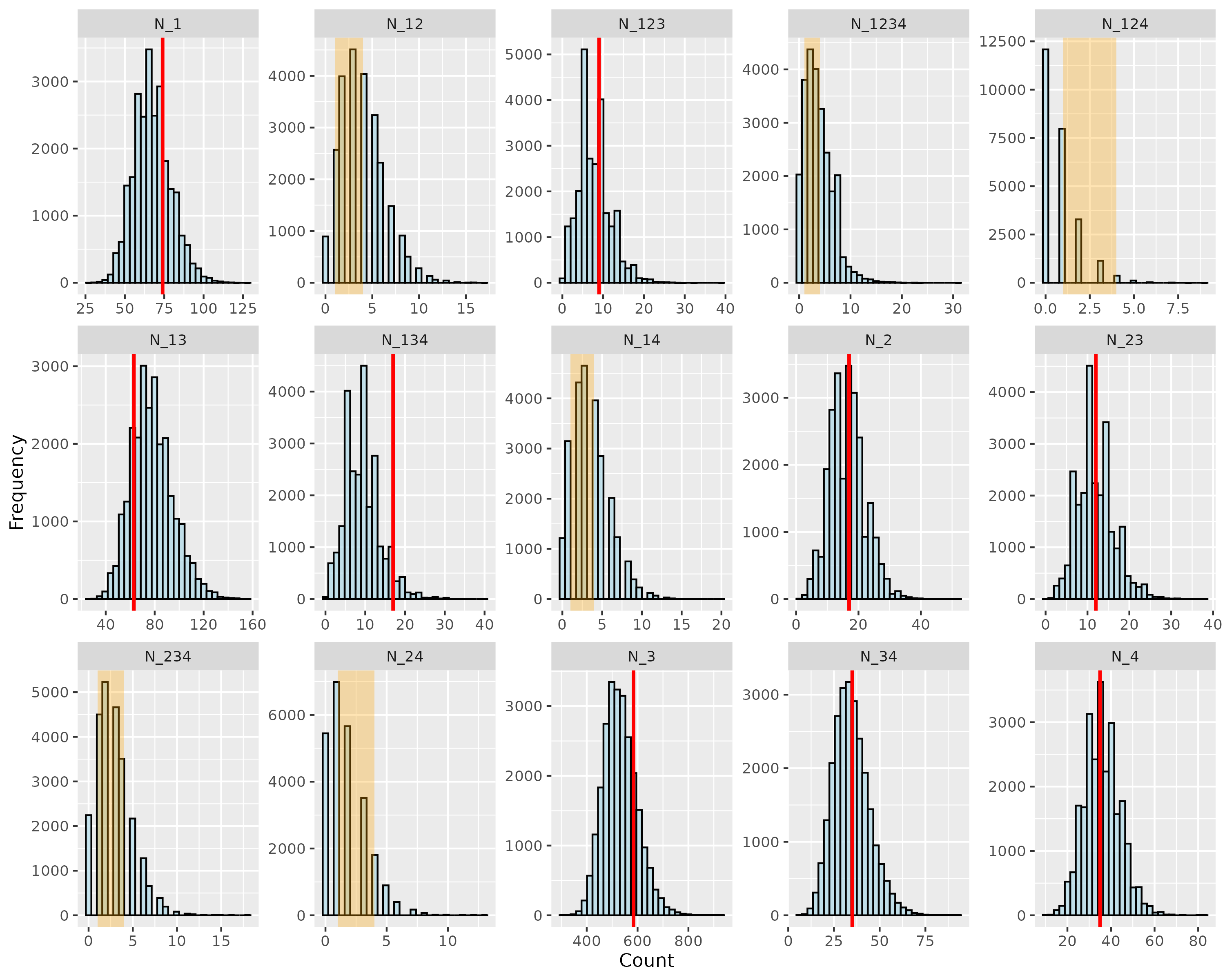}
    \caption{Posterior Predictive Distributions (PPD) for the Female Drug Users dataset. Red vertical lines indicate exact observed counts. Shaded orange bands indicate cells where the data was censored (counts of 1-4). The histograms show that the neural model correctly predicts counts within the censored intervals.}
    \label{fig:king_ppd}
\end{figure}

We further assess model fit using the posterior predictive distributions shown in Figure~\ref{fig:king_ppd}. In this visualization, the solid red lines represent exact observed counts, while the shaded vertical bands represent the censored interval $[1,4]$. A model is considered to fit the censored data well if the probability mass of the predicted count falls substantially within this shaded region.

Visual inspection confirms that the NPE effectively captures the data structure. For main list exact counts (e.g., $n_1$, ..., $n_4$), the observed values fall near the mode of the predictive histograms. Crucially, for the censored interaction terms (e.g., $n_{12}$, $n_{124}$, etc), the posterior predictive distributions place significant mass within the censored regions (orange bands). This indicates that the neural method has successfully learned to allocate probability to these low-count interactions despite the ambiguity introduced by censoring.

\section{Discussion} \label{sec: Discussion}
In this paper, we proposed a simulation-based inference framework for MSE models. This approach allows for robust inference even when count data are censored or missing, which are typically scenarios that render traditional likelihood-based methods intractable or computationally prohibitive. Through extensive simulation studies, we demonstrated that our neural methods can accurately recover model parameters and population sizes in both fully observed and censored settings. Moreover, we showed that the NPE provides uncertainty quantification that is well-calibrated and often superior to MCMC in cases where the latter fails to converge.

The application of our methods to real-world datasets revealed an important distinction between the two neural approaches. For the Modern Slavery dataset, estimates from NBE, NPE, and MCMC were broadly consistent, though the neural methods favored a slightly more conservative hidden population size due to the regularization effect of the prior on unidentifiable interaction terms.

However, the analysis of the Female Drug Users dataset exposed a significant divergence: the NBE estimated a hidden population of 2,530, nearly double the estimates produced by NPE (1,225) and MCMC (1,371).

The NBE is tasked with learning a direct mapping from the sample space to the parameter space. The divergence observed here indicates that the network struggled to learn an accurate mapping for this specific configuration of sparse data. In cases of extreme sparsity, high censorship or atypical datasets, this inverse mapping becomes highly non-linear and complex. The current NBE architecture potentially lacked the necessary flexibility or capacity to capture this complexity. This issue might be remedied in future work by exploring deeper, more expressive architectures (e.g., Transformers or ResNets) or by refining the prior specification to focus the network’s training on the most relevant regions of the parameter space. Consequently, we recommend that researchers treat the NBE as a rapid, first-pass screening tool. When NBE and NPE estimates diverge significantly, the NPE should be considered the more reliable standard, as it allows for posterior predictive checks (Figure~\ref{fig:king_ppd}) that validate whether the learned distribution plausibly explains the observed data.

This difference appears to be driven by a balance between the intercept and the main list effects. The neural estimators infer larger values for $\beta$s compared to MCMC, implying that the probability of appearing on each list is higher than the MCMC estimates. In Multiple Systems Estimation, higher capture probabilities typically result in lower estimates of the hidden population, as fewer unobserved individuals are required to explain the observed counts. Thus, while all Bayesian methods successfully handle the sparsity in the interaction terms, the neural networks settle on a solution with higher catchability and a consequently more conservative hidden population estimate.

It is informative to contrast the behavior of the NBE across the two datasets. In the Modern Slavery application, NBE acted as a regularizer, shrinking the population estimate relative to MLE. However, in the Female Drug Users application, NBE produced a larger population estimate (driven by a larger intercept $\alpha$). This demonstrates that the neural estimators are data-adaptive; they do not simply bias results in a single direction (e.g., always conservative). Instead, they respond to the specific structure of the sparsity and correlation in the observed lists. In this case, inferring lower catchability (more negative $\beta$ values) which necessitates a higher hidden population to explain the observed counts.

The use of Neural Bayes Estimators for Multiple Systems Estimation presents a significant practical advancement, primarily by addressing the computational and accessibility bottlenecks that often hinder traditional inference methods. The chief advantage of our approach is its amortized nature. Once the initial, computationally intensive training phase is complete, generating posterior estimates for new datasets is nearly instantaneous. This is particularly beneficial for MSE studies where data is collected iteratively, allowing researchers to rapidly update their findings without the significant time investment required to re-run a full MCMC analysis. 

Furthermore, the simulation-based framework inherently circumvents pathological estimation scenarios. For instance, in cases where empty overlaps between lists prevent the existence of a finite maximum likelihood estimate, our methods ensure that a stable and finite estimate is always produced.

However, the methods described in this paper come with some limitations. The estimators are trained using a specific, fixed prior distribution. Consequently, if a researcher's prior beliefs differ from those used in training, they cannot be incorporated into the analysis without retraining an entirely new model from scratch. Furthermore, Neural Bayes Estimators return only the posterior median and 95\% credible interval, making model assessment impossible without being able to generate samples from the posterior predictive distribution. 

Moreover, our current framework focuses on parameter estimation for a single, pre-defined model. It does not address the critical step of model selection, where different assumptions about list dependencies can lead to significantly varied hidden population size estimates \citep{Silverman2020, King21}.

Finally, the entire framework's accuracy is contingent on the quality of the simulations used for training. If the training data fails to represent the full spectrum of real-world scenarios, the estimator's performance on unseen data will invariably be poor. Finally, the network itself operates as a `black box'. While we have shown that the estimators can approximate Bayesian posterior quantities with high accuracy, its internal decision-making process is not transparent, which contrasts sharply with the interpretable, step-by-step logic of an MCMC sampler.

For future work, a significant improvement would be to overcome the model's current rigidity. We propose developing a conditional Neural Bayes Estimator that accepts additional inputs alongside the data. These inputs could specify prior hyperparameters or even define the model structure, such as which interaction terms to include. For example, in \cite{JMLR:v25:23-1134} the authors explicitly include the censoring value as an input and demonstrate that the performance in comparable. This would create a single, highly flexible estimator that adapts to a researcher's specific beliefs and hypotheses at inference time, removing the need to retrain a new model for every unique analysis.

\section*{Code, materials, and data availability}

All code and data used in this research can be found in the \url{https://github.com/HiddenHarmsHub/npe-mse-censored-missing} repository.

\section*{Acknowledgments}
This work was supported by a UKRI Future Leaders Fellowship [MR/X034992/1]. The computations described in this paper were performed using the University of Birmingham's BlueBEAR HPC service, which provides a High Performance Computing service to the University's research community. See http://www.birmingham.ac.uk/bear for more details.

\bibliographystyle{apalike}
\bibliography{bibliography}

@article{Luchenski2025,
  title = {Estimating the scale of hospital admissions for people experiencing homelessness in {E}ngland: a population-based multiple systems estimation study using national Hospital Episode Statistics},
  volume = {3},
  ISSN = {2753-4294},
  DOI = {10.1136/bmjph-2025-002978},
  number = {2},
  journal = {BMJ Public Health},
  publisher = {BMJ},
  author = {Luchenski,  Serena April and B\"{o}hning,  Dankmar and Aldridge,  Robert and Stevenson,  Fiona and Tariq,  Shema and Hayward,  Andrew C},
  year = {2025},
  month = jul,
  pages = {e002978}
}

@article{Bird2018,
  title = {Multiple Systems Estimation (or Capture-Recapture Estimation) to Inform Public Policy},
  volume = {5},
  ISSN = {2326-831X},
  DOI = {10.1146/annurev-statistics-031017-100641},
  number = {1},
  journal = {Annual Review of Statistics and Its Application},
  publisher = {Annual Reviews},
  author = {Bird,  Sheila M. and King,  Ruth},
  year = {2018},
  month = mar,
  pages = {95–118}
}

@article{King2008,
  title = {Estimating current injectors in {S}cotland and their drug-related death rate by sex, region and age-group via {B}ayesian capture-recapture methods},
  volume = {18},
  ISSN = {1477-0334},
  DOI = {10.1177/0962280208094701},
  number = {4},
  journal = {Statistical Methods in Medical Research},
  publisher = {SAGE Publications},
  author = {King, Ruth and Bird, Sheila M and Hay, Gordon and Hutchinson, Sharon J},
  year = {2008},
  month = nov,
  pages = {341–359}
}

@Article{SainsburyDale2023,
    title = {Likelihood-Free Parameter Estimation with Neural {B}ayes
      Estimators},
    author = {Matthew Sainsbury-Dale and Andrew Zammit-Mangion and
      Raphael Huser},
    journal = {The American Statistician},
    year = {2024},
    volume = {78},
    pages = {1--14},
    doi = {10.1080/00031305.2023.2249522},
  }

@article{Chan2020,
  title = {Multiple Systems Estimation for Sparse Capture Data: Inferential Challenges When There Are Nonoverlapping Lists},
  volume = {116},
  ISSN = {1537-274X},
  DOI = {10.1080/01621459.2019.1708748},
  number = {535},
  journal = {Journal of the American Statistical Association},
  publisher = {Informa UK Limited},
  author = {Chan,  Lax and Silverman,  Bernard W. and Vincent,  Kyle},
  year = {2020},
  month = feb,
  pages = {1297–1306}
}

@inbook{King21,
  author    = {King, R. and de Rivera Ortega, O. R. and McCrea, R.},
  title     = "Multiple Systems Estimation in the Presence of Censored Cells",
  chapter   = "Book of Short Papers, SIS 2021",
  publisher = "Pearson",
  year      = "2021"
}

@article{Silverman2020,
  title = {Multiple-Systems Analysis for the Quantification of Modern Slavery: Classical and {B}ayesian Approaches},
  volume = {183},
  ISSN = {1467-985X},
  DOI = {10.1111/rssa.12505},
  number = {3},
  journal = {Journal of the Royal Statistical Society Series A: Statistics in Society},
  publisher = {Oxford University Press (OUP)},
  author = {Silverman,  Bernard W.},
  year = {2020},
  month = mar,
  pages = {691–736}
}

@ARTICLE{Gelman1992,
  title     = "Inference from iterative simulation using multiple sequences",
  author    = "Gelman, Andrew and Rubin, Donald B",
  journal   = "Stat. Sci.",
  publisher = "Institute of Mathematical Statistics",
  volume    =  7,
  number    =  4,
  pages     = "457--472",
  month     =  nov,
  year      =  1992
}

@article{TuringJulia,
author = {Fjelde, Tor Erlend and Xu, Kai and Widmann, David and Tarek, Mohamed and Pfiffer, Cameron and Trapp, Martin and Axen, Seth D. and Sun, Xianda and Hauru, Markus and Yong, Penelope and Tebbutt, Will and Ghahramani, Zoubin and Ge, Hong},
title = {Turing.jl: a general-purpose probabilistic programming language},
year = {2025},
publisher = {Association for Computing Machinery},
address = {New York, NY, USA},
doi = {10.1145/3711897},
journal = {ACM Trans. Probab. Mach. Learn.},
month = feb,
}

@article{hoffman2011_NUTS,
author = {Homan, Matthew D. and Gelman, Andrew},
title = {The {N}o-{U}-turn sampler: adaptively setting path lengths in {H}amiltonian {M}onte {C}arlo},
year = {2014},
issue_date = {January 2014},
publisher = {JMLR.org},
volume = {15},
number = {1},
issn = {1532-4435},
journal = {J. Mach. Learn. Res.},
month = jan,
pages = {1593–1623},
numpages = {31}
}

@article{JMLR:v25:23-1134,
  author  = {Jordan Richards and Matthew Sainsbury-Dale and Andrew Zammit-Mangion and Rapha{{\"e}}l Huser},
  title   = {Neural {B}ayes estimators for censored inference with peaks-over-threshold models},
  journal = {Journal of Machine Learning Research},
  year    = {2024},
  volume  = {25},
  number  = {390},
  pages   = {1--49},
  url     = {http://jmlr.org/papers/v25/23-1134.html}
}

@article{SainsburyDale2025,
  title = {Neural {B}ayes Estimators for Irregular Spatial Data Using Graph Neural Networks},
  volume = {34},
  ISSN = {1537-2715},
  DOI = {10.1080/10618600.2024.2433671},
  number = {3},
  journal = {Journal of Computational and Graphical Statistics},
  publisher = {Informa UK Limited},
  author = {Sainsbury-Dale,  Matthew and Zammit-Mangion,  Andrew and Richards,  Jordan and Huser,  Raphaël},
  year = {2025},
  month = jan,
  pages = {1153–1168}
}

@article{LittleCalibration,
  title = {Calibrated {B}ayes, for Statistics in General, and Missing Data in Particular},
  volume = {26},
  ISSN = {0883-4237},
  DOI = {10.1214/10-sts318},
  number = {2},
  journal = {Statistical Science},
  publisher = {Institute of Mathematical Statistics},
  author = {Little,  Roderick},
  year = {2011},
  month = may 
}

@article{Kavianpour2022,
  title = {Next-Generation Capabilities in Trusted Research Environments: Interview Study},
  volume = {24},
  ISSN = {1438-8871},
  DOI = {10.2196/33720},
  number = {9},
  journal = {Journal of Medical Internet Research},
  publisher = {JMIR Publications Inc.},
  author = {Kavianpour, Sanaz and Sutherland, James and Mansouri-Benssassi,  Esma and Coull, Natalie and Jefferson, Emily},
  year = {2022},
  month = sep,
  pages = {e33720}
}

\end{document}